\documentclass[onecolumn,showpacs,preprintnumbers,amsmath,amssymb,superscriptaddress]{revtex4}
\usepackage{graphics,psfrag}
\usepackage{rotating}
\usepackage[usenames]{color}
\usepackage{amsmath}
\usepackage{bbm}
\usepackage{ulem} 
\newcommand{\be}{\begin{eqnarray}}
\newcommand{\ee}{\end{eqnarray}}

\newcommand{\non}{\nonumber\\}
\newcommand{\ave}[1]{\left\langle #1 \right\rangle}

\begin{document}

\title{Chiral dynamics in the $\gamma p \to \pi^0 \eta p$ and
$\gamma p \to \pi^0 K^0 \Sigma^+$ reactions}

\author{M. \surname{D\"oring}}
\email{doering@ific.uv.es}
\affiliation{Departamento de F\'{\i}sica Te\'orica and IFIC,
Centro Mixto Universidad de Valencia-CSIC,\\
Institutos de
Investigaci\'on de Paterna, Aptd. 22085, 46071 Valencia, Spain}
\author{E. \surname{Oset}}
\email{oset@ific.uv.es}
\affiliation{Departamento de F\'{\i}sica Te\'orica and IFIC,
Centro Mixto Universidad de Valencia-CSIC,\\
Institutos de
Investigaci\'on de Paterna, Aptd. 22085, 46071 Valencia, Spain}
\author{D. \surname{Strottman}}
\email{dds@ific.uv.es}
\affiliation{Departamento de F\'{\i}sica Te\'orica and IFIC,
Centro Mixto Universidad de Valencia-CSIC,\\
Institutos de
Investigaci\'on de Paterna, Aptd. 22085, 46071 Valencia, Spain}
\affiliation{Theoretical Division, Los Alamos National Laboratory,
Los Alamos, NM
87545}

\begin{abstract}
Using a chiral unitary approach for meson-baryon scattering in the 
strangeness zero
sector, where the $N^*(1535)S_{11}$ resonance is dynamically generated, we study the
reactions $\gamma p \to \pi^0 \eta p$ and $\gamma p \to \pi^0 K^0 \Sigma^+$ at
photon energies at which the final states are produced close to 
threshold. Among
several reaction mechanisms, we find the most important is the 
excitation of the
$\Delta^*(1700)D_{33}$ state which subsequently decays into a pseudoscalar 
meson and a
baryon belonging to the $\Delta(1232)P_{33}$ decuplet.  Hence, the reaction 
provides useful
information with which to test current theories of the dynamical 
generation of the
low-lying $3/2^-$ states. The first reaction is shown to lead to sizable cross
sections and the $N^*(1535)S_{11}$ resonance shape is seen clearly in the $\eta p$
invariant mass distribution. The same dynamical model is shown to lead to much
smaller cross sections at low energies in the second reaction. 
Predictions are made
for cross sections and invariant mass distributions which can be compared with
ongoing experiments at ELSA.
\end{abstract}
\maketitle
\section{Introduction}
The unitary extensions of chiral perturbation theory $U\chi PT$ have 
brought new
light in the study of the meson-baryon interaction and have shown 
that some well
known resonances qualify as dynamically generated, or in simpler 
words, they are
quasibound states of a meson and a baryon, the properties of which 
are described in
terms of chiral Lagrangians. After early studies in this direction 
explaining the
$\Lambda (1405)S_{01}$ and the $N^*(1535)S_{11}$ as dynamically generated resonances
\cite{Kaiser:1995cy,Kaiser:1996js,kaon,Nacher:1999vg,Oller:2000fj}, 
more systematic
studies have shown that there are two octets and one singlet of 
resonances from the
interaction of the octet of pseudoscalar mesons with the octet of 
stable baryons
\cite{Jido:2003cb,Garcia-Recio:2003ks}. The $N^*(1535)S_{11}$ belongs to 
one of these two
octets and plays an important role in the $\pi N$ interaction with its coupled
channels $ \eta N$, $K \Lambda$ and $K \Sigma$ \cite{Inoue:2001ip}. 
In spite of the
success of the chiral unitary approach in dealing with the 
meson-baryon interaction
in these channels, the fact that the quantum numbers of the $N^*(1535)S_{11}$ are
compatible with a standard three constituent quark structure and that 
its mass is
roughly obtained in many standard quark models 
\cite{Isgur:1978xj,Capstick:1993kb}, or
recent lattice gauge calculations \cite{Chiu:2005zc}, has as a 
consequence that the
case for the $N^*(1535)S_{11}$ to be described as a dynamically generated 
resonance appears
less clean than that of the $\Lambda (1405)S_{01}$ where both quark models 
and lattice
calculations have shown systematic 
difficulties\cite{Nakajima:2001js}. Ultimately, it
will be the ability of the models to describe different experiments 
in which the
resonances are produced that will settle the issue of what represents 
Nature better
at a certain energy scale.  A detailed description of many such 
experiments has been
discussed in
\cite{review}.

A good example is 
found in the
recent experiment on photoproduction of  the $\Lambda (1405)S_{01}$ resonance in the $\gamma 
p \to K^+ \pi
\Sigma$ reaction \cite{Ahn:2003mv}, where theoretical predictions 
using the chiral
unitary approach had been done previously \cite{Nacher:1998mi}.  In 
the present paper
we adopt and extend the ideas of \cite{Nacher:1998mi} and study the 
analogous reaction
$\gamma p \to \pi^0 \eta p$ where the $\eta p $ final state can form 
the $N^*(1535)S_{11}$
resonance. This reaction is currently being analyzed at ELSA\cite{Metag}

Some of the reaction mechanisms in our model are described as a 
two-step process: In
the initial photoproduction, two mesons are generated, one of which 
is the final
$\pi^0$. The final state interaction of the other meson with the proton is then
responsible for the $\eta$ production. For this interaction chiral Lagrangians in $SU(3)$
representation involving only mesons and baryons are used. In addition, the
contributions from explicit baryonic resonance exchange such as $\Delta(1232)P_{33}$,
$N^*(1520)D_{13}$, and $\Delta^*(1700)D_{33}$, which have been 
found essential
for the two meson photoproduction, e.g., in the Valencia model
\cite{Nacher:2000eq,GomezTejedor:1993bq,GomezTejedor:1995pe},  will be included. The 
$\Delta^*(1700)D_{33}$
resonance, as recent studies show
\cite{Kolomeitsev:2003kt,Sarkar:2004jh}, qualifies as dynamically generated through the interaction 
of the $0^-$
meson octet and the $3/2^+$ baryon decuplet . In this picture it is possible
\cite{Sarkar:2004jh} to obtain the coupling of the $\Delta^*(1700)D_{33}$ to the
$\eta\Delta(1232)P_{33}$ and $ K \Sigma^*(1385)P_{13}$ for which experimental 
information does
not yet exist.

In Sec. \ref{sec_pinetan} the model for the dynamical generation of the
$N^*(1535)S_{11}$ resonance in $\pi N\to \pi N$ scattering will be briefly 
reviewed with
special emphasis on the $\pi N\to \eta N$ transition. Subsequently, 
we study the
one-meson photoproduction $\gamma p\to\eta p$ in Sec. 
\ref{sec_onemeson}. This allows
for a simultaneous description of the existing data for these three different
reactions within the chiral model. In Sec. \ref{sec_twofinal} we 
predict observables
for the photoproduction of $\pi^0\eta p$ in the final state.

At the same time we also study the $\gamma p \to \pi^0 K^0 \Sigma^+$ 
reaction and
make predictions for its cross section, taking advantage of the fact 
that it appears
naturally within the coupled channels formalism of the $\gamma p \to 
\pi^0 \eta p$
reaction and leads to a further test of consistency of the ideas explored here.

In this section we have referred to all resonances that will enter the evaluation 
of our amplitudes. In what follows for shortness of notation we will omit
the description in terms of $L_{2I,2J}$.

\section{The $N^*(1535)$ in meson-baryon scattering}
\label{sec_pinetan}
Before turning to the photoproduction reactions of the next sections, 
let us recall
the properties of the $N^*(1535)$ in the meson-baryon sector, where 
this resonance
shows up clearly in the spin isospin $(S=1/2, I=1/2)$ channel. In the 
past, this
resonance has been proposed to be dynamically generated
\cite{Kaiser:1995cy,Kaiser:1996js,Nacher:1999vg,Inoue:2001ip} rather 
than being a
genuine three-quark state. The model of Ref. \cite{Inoue:2001ip} provides 
an accurate
description of the elastic and quasielastic $\pi N$ scattering 
in the $S_{11}$
channel. Within the coupled channel approach in the $SU(3)$ 
representation of Ref.
\cite{Inoue:2001ip}, not only the $\pi N$ final state is accessible, but also
$K\Sigma$, $K\Lambda$, and $\eta N$ in a natural way.

In the case of the present reactions, we are interested in the $\eta 
p$ interaction
which will manifest the $N^\star(1535)$ resonant character. This 
interaction was
studied in detail in Ref. \cite{Inoue:2001ip} for the charge 
$Q=0$, strangeness
zero sector.  In the present study we work in the charge $Q=+1$ 
sector, which requires
the simultaneous consideration of the coupled channels
\be
\pi^0 p,\; \pi^+n,\; \eta p,\;K^+\Sigma^0,\;K^+\Lambda,\;K^0\Sigma^+\; .
\label{channels}
\ee
We will subsequently refer to these channels as one through six in 
the order given
above.  In this section we derive the necessary modifications of the 
coupled channels
in the $Q=+1$ sector and briefly review the basic formalism. The theoretical
framework of the photoproduction mechanisms is found in subsequent sections.

We thus begin with the lowest order chiral Lagrangian for the meson-baryon
interaction
\cite{ulf,ecker}
\be {\cal L}_1^{(B)}& =& \ave{ \bar{B} i \gamma^{\mu} \nabla_{\mu} B} - M_B
\ave{\bar{B} B}
\non &&+\frac{1}{2} D \ave{\bar{B} \gamma^{\mu} \gamma_5 \left\{ u_{\mu}, B
\right\} } + \frac{1}{2} F \ave{\bar{B} \gamma^{\mu} \gamma_5 [u_{\mu}, B]}
\label{pinlagrangian}
\ee where the symbol $\ave{}$ denotes the trace of SU(3) matrices and
\be
\nabla_{\mu} B &=& \partial_{\mu} B + [\Gamma_{\mu}, B], \non
\Gamma_{\mu}& =& \frac{1}{2} (u^+ \partial_{\mu} u + u \partial_{\mu} 
u^+), \non
U&=& u^2 = {\rm exp} (i \sqrt{2} \Phi / f) ,
\non u_{\mu}& =& i u ^+ \partial_{\mu} U u^+
\ee with $\Phi$ and $B$ the usual $3\times 3$ $SU(3)$ matrices of the 
fields for
the meson octet of the pion and the baryon octet of the nucleon, respectively
\cite{ulf}. The term with the covariant derivative $\nabla_\mu$ in Eq.
(\ref{pinlagrangian}) generates the Weinberg-Tomozawa interaction and  leads to
the lowest order transition amplitude
\be V_{ij}=-C_{ij}\;\frac{1}{4f_i f_j}\;\overline{u}(p')\gamma^\mu
u(p)(k_\mu+k_\mu')
\label{kernel}
\ee where $p,p'$ ($k,k'$) are the initial and final momenta of the  baryons
(mesons). The coefficients $C_{ij}$ are $SU(3)$ factors which one obtains from
the Lagrangian, and the $f_i$ are the $\pi$, $\eta$, $K$ decay constants
\cite{gasser}. The
$C_{ij}$ coefficients for the channels with charge +1 are shown in Tab.
\ref{tab:cijs}.
\begin{table}
\caption{$C_{ij}$ coefficients for the six channels.  The matrix is symmetric.}
\begin{center}
\begin{tabular*}{0.7\textwidth}{@{\extracolsep{\fill}}lllllll}
\hline\hline &$\pi^0 p$&$\pi^+ n$&$\eta
p$&$K^+\Sigma^0$&$K^+\Lambda$&$K^0\Sigma^+$
\\
\hline
\rule[-4mm]{0mm}{10mm}$\pi^0 
p$&$0$&$\sqrt{2}$&$0$&$-\frac{1}{2}$&$-\frac{\sqrt{3}}{2}$&$\frac{1}{\sqrt{2}}$
\\
\rule[-4mm]{0mm}{9mm}$\pi^+
n$&&$1$&$0$&$\frac{1}{\sqrt{2}}$&$-\sqrt{\frac{3}{2}}$&$0$
\\
\rule[-4mm]{0mm}{9mm}$\eta
p$&&&$0$&$-\frac{\sqrt{3}}{2}$&$-\frac{3}{2}$&$-\sqrt{\frac{3}{2}}$
\\
\rule[-4mm]{0mm}{9mm}$K^+\Sigma^0$&&&&$0$&$0$&$\sqrt{2}$
\\
\rule[-4mm]{0mm}{9mm}$K^+\Lambda$&&&&&$0$&$0$
\\
\rule[-4mm]{0mm}{9mm}$K^0\Sigma^+$&&&&&&$1$
\\
\hline\hline
\end{tabular*}
\end{center}
\label{tab:cijs}
\end{table}
The amplitudes after unitarization are given in matrix form 
\cite{Inoue:2001ip} by
means of the Bethe-Salpeter equation
\be T(\sqrt{s})=\left[1-V(\sqrt{s})G(\sqrt{s})\right]^{-1} V(\sqrt{s})
\label{bse}
\ee
with $V$ obtained from Eq. (\ref{kernel}).  We are only interested in 
the s-wave
meson-baryon interaction to generate the $S_{11}$ amplitude; the 
projection of $V$
into this partial wave is given in ref. \cite{Inoue:2001ip}, as well 
as $G$ which is
the meson-baryon loop function in dimensional regularization. In the 
following we
denote by $T^{(ij)}$ the matrix elements of $T$ with the channel 
ordering of Eq.
(\ref{channels}).

A second modification of the model of Ref. \cite{Inoue:2001ip} with respect
to other approaches concerns the $\pi N\to \pi\pi N$ channel. This channel was
important to obtain a good description of the $I=3/2$ amplitude but it has only a small influence in the
$I=1/2$ channel. It increases the width by about 10 \% and changes 
the position of the
$N^\star(1535)$ by about 10 MeV. In the charge +1 sector this channel 
can be included
by a change of the potential according to $V_{\pi N, \pi N}\to V_{\pi 
N,\pi N}+\delta
V$ as in Ref. \cite{Doring:2004kt} and reads:
\be
\delta V(\pi^0 p\to \pi^0
p)&=&\left[\left(-\frac{\sqrt{2}}{3}\;v_{31}-\frac{1}{3\sqrt{2}}\;v_{11}\right)^2+\left(\frac{1}{3}\;v_{31}-\frac{1}{3}\;v_{11}\right)^2\right]G_{\pi\pi
N}\non
\delta V(\pi^0 p\to \pi^+
n)&=&\Bigg[\left(-\frac{\sqrt{2}}{3}\;v_{31}-\frac{1}{3\sqrt{2}}\;v_{11}\right)\left(\frac{1}{3}\;v_{31}-\frac{1}{3}\;v_{11}\right)\non
&&+\left(\frac{1}{3}\;v_{31}-\frac{1}{3}\;v_{11}\right)\left(-\frac{1}{3\sqrt{2}}\;v_{31}-\frac{\sqrt{2}}{3}\;v_{11}\right)\Bigg]G_{\pi\pi
N}\non
\delta V(\pi^+ n\to \pi^+
n)&=&\left[\left(\frac{1}{3}\;v_{31}-\frac{1}{3}\;v_{11}\right)^2+\left(-\frac{1}{3\sqrt{2}}\;v_{31}-\frac{\sqrt{2}}{3}\;v_{11}\right)^2\right]G_{\pi\pi
N}
\label{deltavs}
\ee with the isospin classification and conventions as in Ref.
\cite{Inoue:2001ip}; $G_{\pi\pi N}$ being the $\pi\pi N$ loop function that
incorporates the two-pion relative momentum squared.  Analytic expressions for
$v_{11}$ and $v_{31}$ are found in  Ref. \cite{Inoue:2001ip}.

\begin{figure}
\includegraphics[width=15cm]{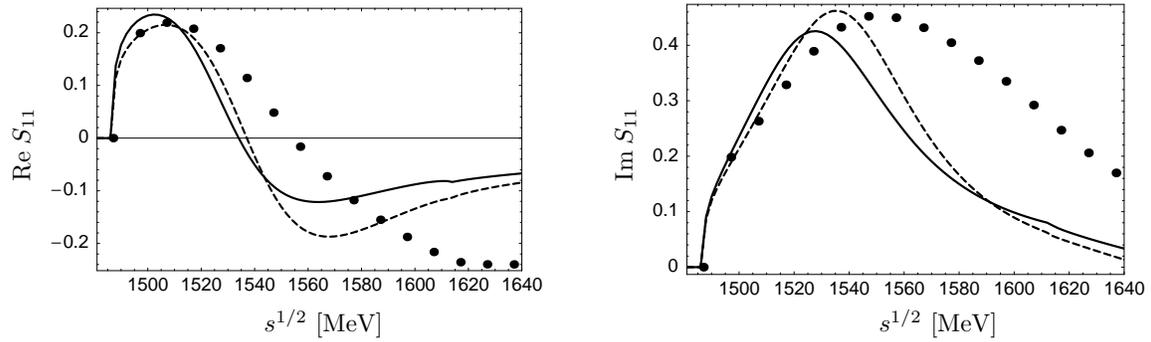}
\caption{The $S_{11}$ partial wave in $\pi N\to \eta N$. Dots: Analysis from
Ref. \cite{Arndt:2003if}. Solid line: full model from Ref. \cite{Inoue:2001ip}.
Dashed line: model from Ref. \cite{Inoue:2001ip} without $t$ channel vector
exchange and $\pi\pi N$ channel.}
\label{fig:s11pinetan}
\end{figure}

The $\pi N\to  \eta N$ production cross section has been calculated in Ref.
\cite{Inoue:2001ip} and was found to be quantitatively correct at the 
peak position,
although somewhat too narrow at higher energies. The question is 
whether this is due
to higher partial waves that enter at larger energies and are not part of the
calculation, or due to a too narrow $N^*(1535)$ of the model.  This can now be
answered because an $S_{11}$ partial wave analysis has become available
\cite{Arndt:2003if}. In Fig. \ref{fig:s11pinetan} this analysis is 
compared to the
model of Ref. \cite{Inoue:2001ip}. With the solid line, the full 
model is indicated,
and with the dashed line the model before introducing the vector 
exchange in the
$t$-channel and the $\pi N\to \pi\pi N$ channel (details in Ref. 
\cite{Inoue:2001ip}).
We will refer to this second one as
a ''reduced'' model of  ref. \cite{Inoue:2001ip} in what follows.
Although we prefer the full model, as form factors and
$\pi\pi N$ production certainly play an important role, we take the differences
between the models in this work as an indication of the theoretical 
uncertainties.

In Fig. \ref{fig:s11pinetan}, the energies close to threshold and in 
particular the
strength are well described by the dynamically generated resonance. 
The position of
the resonance in the analysis \cite{Arndt:2003if} is at slightly higher energies than 
predicted by the
model and the width is considerably larger. This might be due to the 
contribution of
the $N^*(1650)$ resonance which is near the $N^*(1535)$ in the 
$S_{11}$ channel and
has been found to contribute to the reaction in other work 
\cite{Gasparyan:2003fp}.
Note, however, that in the same reference the total cross section 
above $s^{1/2}$
around 1650 MeV is dominated by heavier resonances from other partial 
waves such as
the $P_{13}(1720)$ and $D_{13}(1520)$. It is also worth noting that 
in some variants
of the chiral models with additional input to the one used here, one 
can account for
the $N^*(1650)$ contribution to the $S_{11}$ amplitude\cite{Nieves}.

In the present approach we restrict ourselves to the model for the 
$N^*(1535)$ from
Ref. \cite{Inoue:2001ip} as the behavior near the $\eta p$ threshold is well
described in that work including the strength at the maximum of the 
cross section.

\subsection{Single meson photoproduction}
\label{sec_onemeson}
\begin{figure}
\includegraphics[width=15cm]{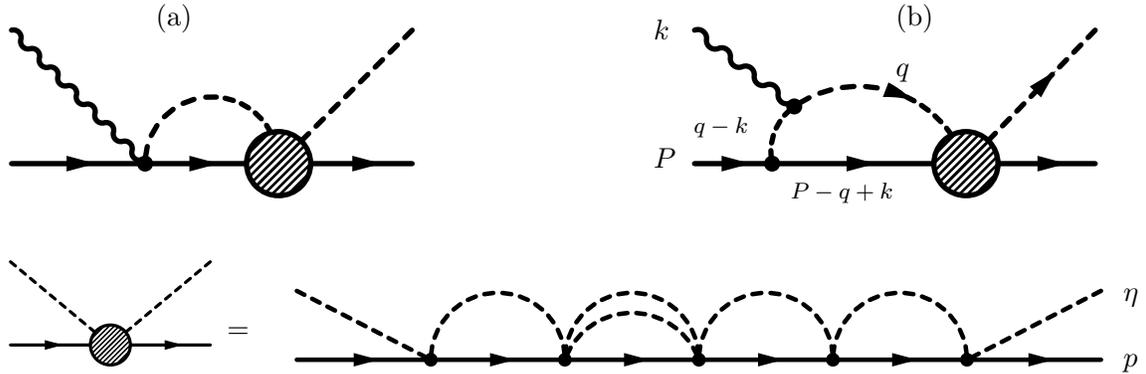}
\caption{Photoproduction of $\eta p$ via the $N^*(1535)$ resonance (gray blob).
Kroll-Ruderman term (a) and the meson pole term (b).}
\label{fig:feynetap}
\end{figure}

In the previous section we have seen that the dynamically generated 
$N^*(1535)$
resonance provides the correct strength in the $\pi N\to  \eta N$ 
transition at low
energies. Here, we test the model for the reaction $\gamma p\to \eta 
p$ with the
basic photoproduction mechanisms plotted in Fig. \ref{fig:feynetap}, 
which consist
of the meson pole term and the Kroll-Ruderman term -- included for gauge
invariance -- followed by the rescattering of the intermediate charged meson
described by the model of the last section.
We shall come back to the question of gauge invariance later on in the section 
by  looking at other, subdominant diagrams.

The baryon-baryon-meson (BBM) vertex is given by the chiral Lagrangian
\be {\cal L}_{{\rm BBM}}=\frac{D+F}{2}\ave{\overline{B}\gamma^\mu\gamma_5 u_\mu
B}+\frac{D-F}{2}\ave{\overline{B}\gamma^\mu\gamma_5 B u_\mu}
\label{LMBB}
\ee with the notation from Sec. \ref{sec_pinetan}. The Kroll-Ruderman term is
obtained from this interaction by minimal substitution and the $\gamma MM$
couplings emerge from scalar QED. For the $i$th meson-baryon channel from Eq.
(\ref{channels}), the $T$-matrix elements read
\be
t_{KR}^i(\sqrt{s})&=&-\frac{\sqrt{2}ie}{f_i}\; {\vec \sigma}{\vec
\epsilon}\;\left(a_{\rm KR}^i\;\frac{D+F}{2}+b_{\rm
KR}^i\;\frac{D-F}{2}\right)\;T^{(i3)}(\sqrt{s})\int\limits^\Lambda\frac{d^3{\bf 
q}}{(2\pi)^3}
\frac{M}{2\omega({\bf q})E({\bf q})}\frac{1}{\sqrt{s}-E({\bf 
q})-\omega({\bf q})+i\epsilon}
\non
t_{MP}^i(\sqrt{s})&=&-\frac{\sqrt{2}}{f_i}\;{\vec
\sigma}{\vec \epsilon}\left(a_{\rm BBM}^i\;\frac{D+F}{2}+b_{\rm
BBM}^i\;\frac{D-F}{2}\right)(-iec_{\gamma MM}^i)T^{(i3)}(\sqrt{s})\non
&&\frac{M}{2(2\pi)^2}\int\limits_0^\Lambda
dq\;q^2\int\limits_{-1}^{1}dx\;\frac{q^2(1-x^2)}{E(q)}\frac{1}{\sqrt{s}-\omega-E(q)+i\epsilon}\;\frac{1}{\sqrt{s}-\omega'-k-E(q)+i\epsilon}\non
&&\frac{1}{\omega\omega'}\;\frac{1}{k-\omega-\omega'+i\epsilon}\;\frac{1}{k+\omega+\omega'}\;\left[k\omega'+(E(q)-\sqrt{s})(\omega+\omega')+(\omega+\omega')^2\right]
\label{krmp}
\ee
for the Kroll-Ruderman term and the meson pole, respectively. The 
amplitudes $T^{(i
3)} (\sqrt{s}\:)$ are the strong transition amplitudes from channel 
$i$ to the $\eta
p$ channel, following the ordering of table 1. In Eq. (\ref{krmp}) 
and throughout this study we
use the Coulomb gauge  ($\epsilon^0=0, {\vec \epsilon\cdot}{\bf 
k}=0$, with ${\bf k}$
the photon three-momentum). The assignment of momenta in Eq. 
(\ref{krmp}) is given
according to Fig. \ref{fig:feynetap}, $\omega=\sqrt{q^2+m_\pi^2}$,
$\omega'=\sqrt{q^2+k^2-2qkx+m_\pi^2}$, $E(q)$ the baryon energy, 
$\sqrt{s}=P^0+k^0$
and $G$ being the meson-baryon loop function according to Ref. 
\cite{Inoue:2001ip}.
The coefficients $a,b,c$ are given in Table \ref{tab:krmb}.
\begin{table}
\caption{Isospin coefficients for the Kroll-Ruderman term ($a_{\rm
KR}^i,\;b_{\rm KR}^i$), BBM vertex ($a_{\rm BBM}^i,\;b_{\rm BBM}^i$), and
$\gamma MM$ vertex ($c_{\gamma MM}^i$).}
\begin{center}
\begin{tabular*}{0.7\textwidth}{@{\extracolsep{\fill}}lllllll}
\hline\hline
&$\pi^0 p$&$\pi^+ n$&$\eta p$&$K^+\Sigma^0$&$K^+\Lambda$&$K^0\Sigma^+$
\\
\hline
\rule[-4mm]{0mm}{10mm}$a_{\rm KR}^i$&$0$&$-1$&$0$&$0$&$\sqrt{\frac{2}{3}}$&$0$
\\
\rule[-4mm]{0mm}{9mm}$b_{\rm 
KR}^i$&$0$&$0$&$0$&$-\frac{1}{\sqrt{2}}$&$-\frac{1}{\sqrt{6}}$&$0$
\\
\rule[-4mm]{0mm}{9mm}$a_{\rm 
BBM}^i$&$\frac{1}{\sqrt{2}}$&$1$&$\frac{1}{\sqrt{6}}$&$0$&$-\sqrt{\frac{2}{3}}$&$0$
\\
\rule[-4mm]{0mm}{9mm}$b_{\rm 
BBM}^i$&$0$&$0$&$-\sqrt{\frac{2}{3}}$&$\frac{1}{\sqrt{2}}$&$\frac{1}{\sqrt{6}}$&$1$
\\
\rule[-4mm]{0mm}{9mm}$c_{\gamma MM}^i$&$0$&$-1$&$0$&$-1$&$-1$&$0$
\\
\hline\hline
\end{tabular*}
\end{center}
\label{tab:krmb}
\end{table}
The cut-off in Eq. (\ref{krmp}) has been chosen $\Lambda=1400$ MeV. With this value, the reduced model 
of the rescattering (see comment below Eq. (\ref{deltavs}))
provides the same
strength as the data \cite{said_solution} at the maximum position
of the total cross section. Once the cut-off has been fixed, we 
continue using this
value for $\Lambda$ in the following sections, for the reduced and full model.

\begin{figure}
\includegraphics[width=11cm]{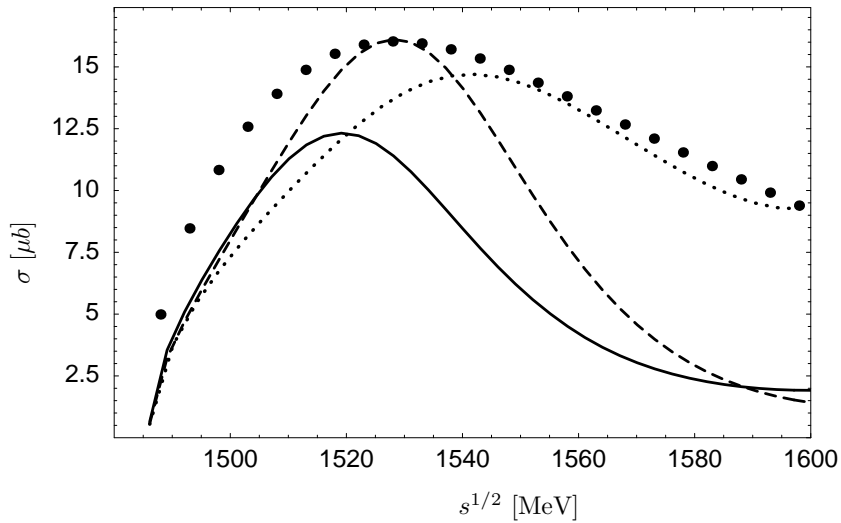}
\caption{Cross section for $\gamma p\to\eta p$. Dots: data from
Ref. \cite{said_solution}. Solid line: Prediction including the full 
model from Ref.
\cite{Inoue:2001ip}. Dashed line: Reduced model from Ref. 
\cite{Inoue:2001ip}. Thin
dotted line: Phenomenological $\pi N\to \eta N$ potential from from Ref.
\cite{Arndt:2003if}.}
\label{fig:csonemeson}
\end{figure}

The amplitudes are unitarized by the coupled channel approach from Ref.
\cite{Inoue:2001ip} in the final state interaction which provides at 
the same time
the $\eta$ production. This is indicated diagrammatically in Fig. 
\ref{fig:feynetap}
with the gray blob. The total amplitude including the rescattering 
part is then given
by
\be
T_{\gamma p\to\eta
p}(\sqrt{s}\:)=\sum_{i=1}^6 t_{KR}^i(\sqrt{s}\:)+t_{MP}^i(\sqrt{s}).
\label{alloneprod}
\ee
The resulting cross section is plotted in Fig. \ref{fig:csonemeson} 
together with the
data compilation from Ref. \cite{said_solution}.
With the solid line, the result including the full model for the $MB\to\eta p$
transition according to Sec. \ref{sec_pinetan} is plotted (dashed line: reduced
model). The diagrams from Fig. \ref{fig:feynetap}, together with the
unitarization, explain quantitatively the one-meson photoproduction 
at low energy
which indicates that these mechanisms should be included in the two-meson
photoproduction reactions of the next section.

Instead of our microscopic description of the $\eta$ production, one 
can also insert
the phenomenological $\pi N\to \eta N$ transition amplitude from Sec.
\ref{sec_pinetan} and Ref. \cite{Arndt:2003if} into the rescattering 
according to
Fig. \ref{fig:feynetap}. The channels $K^+\Sigma^0$ and $K^+\Lambda$ 
in the first loop
play an important role and should be incorporated as initial states in the
$MB\to \eta p$ transition. In this case we include them by replacing 
$T^{(i3)}$ in
Eq. (\ref{alloneprod}) by
\be
T^{(i3)}(\sqrt{s}\:)\to \frac{T_{\rm
ph}^{(23)}(\sqrt{s}\:)}{T^{(23)}(\sqrt{s}\:)}\;T^{(i3)}(\sqrt{s}\:)
\label{phenonepord}
\ee
where $T_{\rm ph}^{(23)}$ is the  phenomenological $S_{11}$ amplitude to the
transition $\pi^+ n\to\eta p$. The prescription of Eq. 
(\ref{phenonepord}) is the
correct procedure for the $\pi^+ p$ channel which is the dominant 
one, and we assume
it to be valid for the other channels. We choose $\Lambda=1400$ MeV for the
cut-off as before.
The cross section is displayed in Fig.
\ref{fig:csonemeson} with the thin dotted line and indeed shows a wider shape.

As we can see in Fig. \ref{fig:csonemeson}, the description of the data is only qualitative.
Given the theoretical uncertainties one should not pretend a better agreement with the data. Yet, in both
theoretical calculations the distribution is too narrow, reflecting most probably the lack of the $N^*(1650)S_{11}$ contribution in the theoretical calculation. The uncertainties of the model for this reaction will be considered later on in the study of the $\gamma p\to \pi^0\eta p$ reaction in order to estimate its theoretical uncertainties.

\vspace*{0.3cm}

At this point we would like to make some general comments concerning basic symmetries 
and the degree to which they are respected in our approach, as for instance chiral symmetry 
or gauge invariance. 

In our approach we are using chiral Lagrangians which are used as the kernel of the Bethe Salpeter
equation and which are chiral symmetric up to mass terms which explicitly break the symmetry. 
The unitarization does not break this symmetry of the underlying theory since it is respected in chiral perturbation theory 
($\chi$PT), and a perfect matching with $\chi$PT to any order can be obtained with the approach that we
use, as shown in Ref. \cite{Oller:2000fj}.

Tests of symmetries can be better done in field theoretical approaches that use, for instance, dimensional regularization for the loops. Although dimensional regularization is used here in the loops for meson baryon scattering, we have preferred to use a cut off for the first loop involving the photon and do some fine tuning to fit the data. Then, we use this cut off 
(which is well within reasonable values) for the other loops that we will find later on. 
The cut off method is also easier and more transparent when dealing with particles with a finite width as it will be our case.
The use of this cut off scheme 
or the dimensional regularization are in practice identical, given the matching between the two loop functions done in Section 2 of 
Appendix A of Ref. \cite{Oller:1998hw}. There, one finds that the dimensional regularization formula and the one with cut off have the same analytical properties (the log-terms) and are numerically equivalent for values of the cut off reasonably larger than the on shell momentum 
of the states of the loop, which is a condition respected in our calculations. By fine tuning the subtraction constant in dimensional regularization, or fine tuning the cut off, one can make the two expressions identical at one energy and practically equal in a wide range of energies, sufficient for studies like the present one. Of particular relevance is the explicit appearance of the log-terms in the cut off scheme which preserve all the analytical properties of the scattering amplitude. 

The equivalence of the schemes would also guarantee that gauge invariance is preserved with the cut off scheme if it is also the case in dimensional regularization. This of course requires that a full set of Feynman diagrams is chosen which guarantees gauge invariance.  At this point we can clearly state that the set of diagrams chosen in Fig. \ref{fig:feynetap} is not gauge invariant. Some terms are missing, which we describe below, and which are omitted because from previous studies we know they are negligible for low energy photons \cite{Lee:1998gt}. Since the energy  of the photon is not so small here, it is worth retaking the discussion which we do below. 

The issue of gauge invariance for pairs of interacting particles has received certain attention \cite{Gross:1987bu,Kvinikhidze:1998xn,vanAntwerpen:1994vh,Haberzettl:1997jg}, but for the purpose of the present paper we can quote directly 
the work of \cite{Borasoy:2005zg} which proves that when using the Bethe Salpeter equation with the kernel of the Weinberg-Tomozawa term, as we do here, gauge invariance is automatically satisfied when the coupling of the photon is made not only to the external legs and vertices, but also to the vertices and intermediate particle propagators of the internal structure of the Bethe-Salpeter equation.

\begin{figure}
\includegraphics[width=13cm]{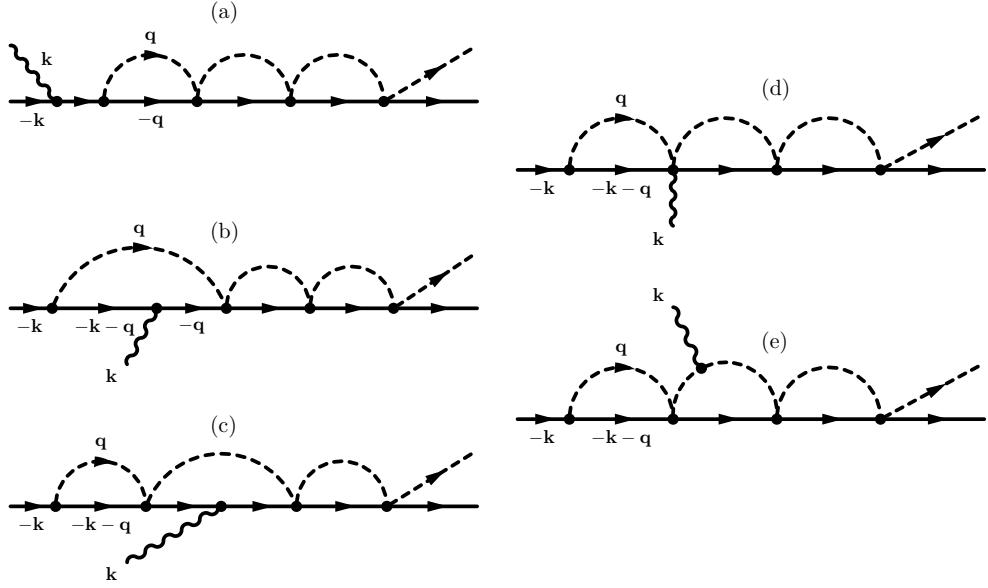}
\caption{Photon coupling besides the diagrams from Fig. \ref{fig:feynetap}. The photon can also couple to the external baryon (a), internal baryon of the first loop (b), and components of the rescattering (c)-(e).}
\label{fig:addsmall}
\end{figure}

A complete set of diagrams fulfilling gauge invariance requires in addition to the diagrams shown in Fig. \ref{fig:feynetap} (a), (b), other diagrams where the photon couples to the baryon lines, vertices, or the internal meson lines from rescattering. We plot such diagrams in Fig. \ref{fig:addsmall}. 
All of them vanish in the heavy baryon approximation. This is easy to see. In diagram (a), Fig. \ref{fig:addsmall}, the first loop to the left (think for the moment about a $\pi^+n$ loop) contains a $p$-wave vertex of the $\boldsymbol{\sigma}\cdot {\bf q}$ type and an $s$-wave vertex, and vanishes in any case. Diagram (b) contains a $p$-wave and an $s$-wave vertex in the loop plus a $\gamma nn$ vertex proportional to $\boldsymbol{\sigma}\times {\bf k}$. In the baryon propagators one momentum is ${\bf q}$ and the other one ${\bf q+k}$ and the integral does not vanish. However, the contribution is of the order $\left(\frac{k}{2M_p}\right)^2$ or 5\%. The term (c) has the same property, a $p$-wave and an $s$-wave  vertex in the first loop to the left, and only the fact that the propagator depends on ${\bf k+q}$ renders a small contribution (remember we are performing a nonrelativistic calculation by taking $\boldsymbol{\sigma}\cdot {\bf q}$ for the Yukawa vertices, but this is more than sufficient for the estimates we do). In diagram (d) the $MMBB\gamma$ vertex is of the type $(\boldsymbol{\sigma}\times {\bf q})\cdot\boldsymbol{\epsilon}$ (see Sec. \ref{sec:no_deltas}), hence once again we have the same situation as before for the first loop to the left. Finally, in diagram (e) the photon is coupled to the internal meson line of the rescattering. In this case both the loop of the photon as well as the first one to the left contain just one $p$-wave coupling and the diagram is doubly suppressed. 

In order to know more precisely how small are the diagrams in our particular case we perform the calculation of one of them, diagram (b), explicitly. By assuming a $\pi^+n$ in the loop to the left in diagram (b), we obtain for the loop
\be
\tilde{t}^{(b)}&=&\frac{\mu_n}{2M}\;\boldsymbol{\sigma}\cdot\boldsymbol{\epsilon}\;e\sqrt{2}\;\frac{D+F}{2f_\pi} \int\frac{d^3{\bf q}}{(2\pi)^3}\non
&&
\frac{1}{2\omega({\bf q})}\frac{M}{E_n({\bf q})}\frac{M}{E_N({\bf k+q})}\frac{1}{\sqrt{s}-\omega({\bf q})-k-E_N({\bf k+q})+i\epsilon}\frac{1}{\sqrt{s}-\omega({\bf q})-k-E_N({\bf q})+i\epsilon}\;{\bf q\cdot k}
\ee
while the equivalent loop function for the Kroll-Ruderman term would be
\be
\tilde{t}^{({\rm KR})}&=&-\boldsymbol{\sigma}\cdot\boldsymbol{\epsilon}\;e\sqrt{2}\;\frac{D+F}{2f_\pi} \int\frac{d^3{\bf q}}{(2\pi)^3}\;\frac{1}{2\omega({\bf q})}\frac{M}{E_n({\bf q})}\frac{1}{\sqrt{s}-\omega({\bf q})-k-E_N({\bf q})+i\epsilon}
\ee
where $\mu_n$ is the neutron magnetic moment. The explicit evaluation of the terms $\tilde{t}^{(b)}$ and $\tilde{t}^{({\rm KR})}$ indicates that when the two terms are added coherently there is a change of  6\% in $|t|^2$ with respect to the Kroll-Rudermann term alone. If we add now the term with $\pi^0 p$ in the intermediate state of the first loop and project over $I=1/2$ to match with $\eta N$ in the final state, the contribution of the magnetic part is proportional to $2\mu_n+\mu_p$ instead of $2\mu_n$ with the $\pi^+n$ state alone, and the contribution becomes of the order of 2\%. The convection term $e({\bf p+p'})\cdot\boldsymbol{\sigma}/(2M)$ ($p,\; p'$ nucleon momenta) of the $\gamma pp$ coupling (not present for the neutron) leads to an equally small contribution. 

The exercise tells us how small is the contribution that vanishes exactly in the heavy baryon limit. Since we do not aim at a precision of better than 20 \% these terms are negligible for us and hence are not further considered. 

\section{Eta pion photoproduction}
\label{sec_twofinal}
Having reviewed the single $\eta$ production in the meson-baryon sector and having
applied the model to the single $\eta$ photo production we turn now to 
the more complex
reaction $\gamma p\to\pi^0\eta p$. The reaction will be discussed in 
three steps: In
the first part, the participating hadrons will be only mesons and 
baryons with their
chiral interaction in $SU(3)$.  In the second part the contributions 
from explicit
baryonic resonances will also be taken into account as they are known 
to play an
important role, e.g., in the two pion photoproduction
\cite{Nacher:2000eq,GomezTejedor:1993bq,GomezTejedor:1995pe}. 
Finally, the decay
channels of the $\Delta^*(1700)$ into $\eta\Delta(1232)$ and $K\Sigma^*(1385)$ 
will be included.

\subsection{Contact interaction and anomalous magnetic moment}
\label{sec:no_deltas} 
One of the important  features of the models 
for reactions that
produce dynamically generated resonances is that  the Lagrangians do 
not involve
explicitly the resonance degrees of freedom. Thus, the  coupling of 
photons and mesons
is due to the more elementary components, in this  case the mesons 
and baryons, which
are the building blocks of the coupled channels  and which lead to 
the resonance
through their interactions.

\begin{figure}
\includegraphics[width=12cm]{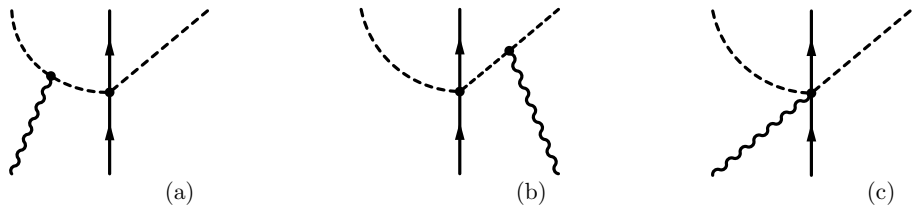}
\caption{Photon interaction with mesons and a baryon. The straight dashed line
symbolizes an outgoing meson and the curved line the meson in a loop 
of the final
state interaction.}
\label{fig:photon_vertex}
\end{figure}

We follow the formalism of Ref. \cite{Nacher:1998mi} for the 
$\gamma p\to
K^+\pi\Sigma$ reaction where the $\Lambda(1405)$ resonance is clearly 
visible in the
$\pi\Sigma$ invariant mass distribution.  The derivative coupling in 
the meson vertex
of Eq. (\ref{kernel})  leads to a $\gamma MMBB$ contact vertex through minimal
coupling, see Fig.  \ref{fig:photon_vertex} (c), and guarantees gauge 
invariance
together with the meson pole terms of Fig. \ref{fig:photon_vertex} (a),(b).

The contact term of Fig. \ref{fig:photon_vertex} (c) is easily 
generated and assuming
the reaction $\gamma M_i B_i\to M_j B_j$ the amplitude is given by
\be 
V_{ij}^{(\gamma)}=C_{ij}\;\frac{e}{4f_if_j}\;(Q_i+Q_j)\overline{u}(p')\gamma^\mu
u(p)\epsilon_\mu \;
\ee
with $Q_i, Q_j$ the meson charges. In the Coulomb gauge this becomes
\be 
V_{ij}^{(\gamma)}=-C_{ij}\;\frac{e}{4f_if_j}\;i\;\frac{\vec{\sigma}\times 
{\bf
q}}{2M_p}\;{\vec \epsilon}\;(Q_i+Q_j)
\label{another_v}
\ee
in the $\gamma p$ CM frame. Since the initial channel $i$ is $\pi^0 
p$, or channel
number 1 in the order of the channels from Sec. \ref{sec_pinetan}, we obtain
\be
V_{1j}^{(\gamma)}=-C_{1j}\;\frac{e}{4f_1f_j}\;i\;\frac{\vec{\sigma}\times
{\bf q}}{2M_p}\;{\vec \epsilon}\; \;Q_j.
\label{gammaBMMB}
\ee
It was shown in Ref. \cite{Nacher:1998mi} that the meson pole terms of Fig.
\ref{fig:photon_vertex} (a), (b) are small compared to the amplitude of Eq.
(\ref{gammaBMMB}) for energies where the final particles are 
relatively close to
threshold, as is the case here, both at the tree level or when the 
photon couples to
the mesons within loops. The coupling of the photon to the baryon 
components  was also
small and will be neglected here, as was done in Ref. \cite{Nacher:1998mi}.

Before we proceed to unitarize the amplitude, it is worth looking at the
structure of Eq. (\ref{gammaBMMB}) which contains the ordinary magnetic moment
of  the proton. It is logical to think that a realistic amplitude 
should contain
also the  anomalous part of the magnetic moment. This is indeed the case if one
considers the effective Lagrangians given in Ref. \cite{stein}
\be {\cal
L}&=&-\;\frac{i}{4M_p}\;b_6^F\ave{\overline{B}\left[S^\mu,S^\nu\right]\left[F^+_{\mu\nu},B\right]}\non
&&-\;\frac{i}{4M_p}\;b_6^D\ave{\overline{B}\left[S^\mu,S^\nu\right]\{F^+_{\mu\nu},B\}}
\label{anola}
\ee
with
\be
F_{\mu\nu}^+&=&-e\left(u^\dagger
QF_{\mu\nu}\;u+uQF_{\mu\nu}\;u^\dagger\right),\non
F_{\mu\nu}&=&\partial_\mu A_\nu-\partial_\nu A_\mu
\label{fmunu}
\ee
with $M_p$ the proton mass and $A_\mu$ the electromagnetic field. The 
operator $Q$ in
Eq. (\ref{fmunu}) is the quark charge matrix $Q={\rm diag}(2,-1,-1)/3$ and
$S^\mu$ is the spin matrix which in the rest frame becomes 
$(0,\vec{\sigma}/2)$. In
Ref.  \cite{Jido:2002yz} the Lagrangians of Eq. (\ref{anola}) were 
used to determine
the magnetic  moment of the $\Lambda(1405)$. In the Coulomb gauge one 
has for an
incoming photon
\be
\left[S^\mu, S^\nu\right]F_{\mu\nu}\to \left(\vec{\sigma}\times {\bf 
q}\right){\vec
\epsilon}
\ee and, thus, the vertex from the Lagrangian of Eq. (\ref{anola}) can 
be written as
\be
{\cal L}\to e\;\frac{\vec{\sigma}\times {\bf q}}{2M_p}\;{\vec
\epsilon}\;\Big(\frac{i}{2}\;b_6^F\ave{\overline{B}\left[
\left(u^\dagger Q u+uQu^\dagger\right),B\right]}\non
+\frac{i}{2}\;b_6^D\ave{\overline{B}\{\left(u^\dagger Q
u+uQu^\dagger\right),B\}}\Big)
\label{anoeval}.
\ee
Expanding the terms up to two meson fields leads to contact vertices 
with the same
structure as Eq. (\ref{another_v}).  Taking $u=1$ in Eq. 
(\ref{anoeval}), and hence
with no meson fields, provides the full magnetic moments of the octet 
of baryons from
where one obtains the values of the coefficients \cite{stein, Jido:2002yz}
\be
b_6^D=2.40, \quad b_6^F=1.82. \nonumber
\ee
It is easy to see \cite{Jido:2002yz} that by setting $b_6^D=0$, 
$b_6^F=1$, one obtains
the ordinary magnetic moments of the baryons without the anomalous 
contribution.
Similarly, taking the same values of $b_6^D$, $b_6^F$ one obtains Eq. 
(\ref{another_v})
for the vertices $\gamma MMBB$. This is easily seen by explicitly 
evaluating the
matrix elements of Eq. (\ref{anoeval}) which lead to the amplitude
\be
-it_{ij}^\gamma=-\;\frac{e}{2M_p}(\vec{\sigma}\times {\bf q}){\vec
\epsilon }\;\frac{1}{2f_if_j}\left[X_{ij}b_6^D+Y_{ij}b_6^F\right]
\label{tij}
\ee
where the coefficients $X_{ij}$ and $Y_{ij}$ are given in Table \ref{tab:xijs}.
\begin{table}
\caption{$X_{1j}$ and $Y_{1j}$ coefficients for the anomalous magnetic
moment.}
\begin{center}
\begin{tabular*}{0.7\textwidth}{@{\extracolsep{\fill}}lllllll}
\hline\hline
&$\pi^0 p$&$\pi^+ n$&$\eta p$&$K^+\Sigma^0$&$K^+\Lambda$&$K^0\Sigma^+$
\\
\hline
\rule[-4mm]{0mm}{10mm}$X_{1j}$&$0$&$\sqrt{2}$&$0$&$\frac{1}{2}$&$-\frac{1}{2\sqrt{3}}$&$0$
\\
\rule[-4mm]{0mm}{10mm}$Y_{1j}$&$0$&$\sqrt{2}$&$0$&$-\frac{1}{2}$&$-\frac{\sqrt{3}}{2}$&$0$
\\
\hline\hline
\end{tabular*}
\end{center}
\label{tab:xijs}
\end{table}
The combination of the $Y_{1j}$ in Table \ref{tab:xijs} and the $C_{1 
j}$ of Table
\ref{tab:cijs} shows  the identity of Eq. (\ref{tij}) and Eq. 
(\ref{gammaBMMB}) for
the case of
$b_6^D=0$, $b_6^F=1$.

\subsubsection{Unitarization}
\begin{figure}
\includegraphics[width=14cm]{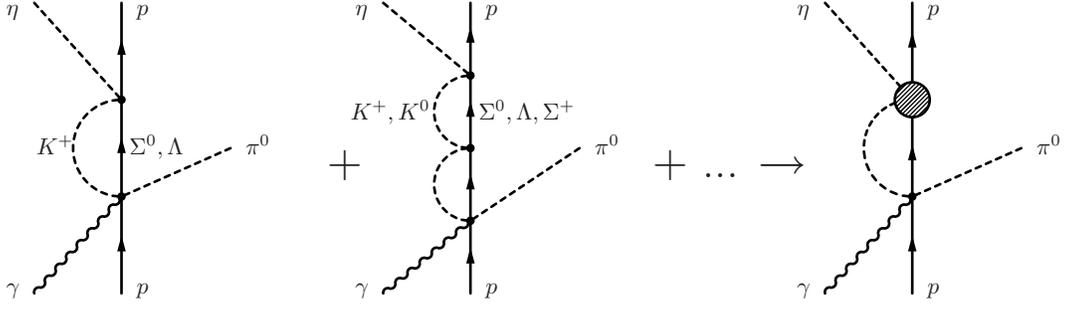}
\caption{Unitarization of the transition amplitude for $\eta p$ production. The
possible states (see table \ref{tab:cijs}) for one and two loops are 
indicated.}
\label{fig:rescattering_diagrams}
\end{figure}

For the amplitude $\gamma p \to \pi^0\eta p$, the first thing  to 
realize is that at
tree level the amplitude is zero with the interactions from Eqs. 
(\ref{gammaBMMB}) and
(\ref{tij}). It is the unitarization and the coupled channel procedure that
renders this amplitude finite and sizable. The unitarization procedure with the
coupled channels allows the intermediate channels with charged mesons of Eq.
(\ref{channels}) to be formed,  even if some of them are not 
physically open. The
scattering of these  states leads finally to $\eta p$. 
Diagrammatically, this is
depicted in Fig.
\ref{fig:rescattering_diagrams}
which implicitly assumes the unitarization is implemented via the use of the
Bethe-Salpeter equation (\ref{bse}) which generates the diagrams of Fig.
\ref{fig:rescattering_diagrams}.

Since $\eta p$ is channel 3 in our list of coupled channels, our 
final amplitude reads
\be T_{\gamma p\to \pi^0\eta p}=-i\sum_j \left(b_6^D X_{1j}+b_6^F
Y_{1j}\right)\frac{e}{4f_1f_j}\;\frac{\vec{\sigma}\times {\bf
q}}{2M_p}\;\vec{\epsilon} \;G_j(z)\;T^{(j3)}(z)\; ,
\label{t13_contact}
\ee
where $G_j$ is the meson-baryon loop function which is obtained in Ref.
\cite{Inoue:2001ip} using dimensional regularization, and the 
$T^{(j3)}$ are the
ordinary scattering matrices of the $\eta p$ and coupled channels 
from Eq. (\ref{bse}).
The invariant kinematical argument $z$ is given by the invariant mass 
$M_I(\eta p)$ of
the $\eta p$ system,
\be
z=M_I
\label{z1}
\ee
or, alternatively,
\be
z=\left(s+m_\pi^2-2\sqrt{s}\;p_\pi^0\right)^{1/2}
\label{z2}
\ee
with $p_\pi^0=({\bf p}_\pi^2+m_\pi^2)^{1/2}$, when the amplitude is 
expressed in terms
of the invariant mass $M_I(\pi^0 p)$ of the $\pi^0p$ system.

One might also question why we do not unitarize the other $\pi^0$ 
with the $\eta$ or
the proton. The reason has to do with the chosen kinematics. By being close to
threshold the $\pi^0$ has a small momentum and is far from the region of the
$a_0(980)$ resonance that could be created interacting with the $\eta$.  The
generation of the $\pi^0 p$ invariant masses in the $\Delta(1232)$ 
region in the phase
space that we investigate is more likely.  However, as one can see 
from Fig. 5 an
extra loop of the $\pi^0$ and $p$ lines produces a $\Delta(1232)$ 
which would involve
an s-wave vertex and a p-wave vertex.  This would vanish in the loop 
integration in
the limit of large baryon masses.  Later, we shall consider other 
diagrams in which
the $\Delta(1232)$ is explicitly produced.

\subsection{Kroll-Ruderman and meson pole term}
\label{sec:kr_mepo}
\begin{figure}
\includegraphics[width=12cm]{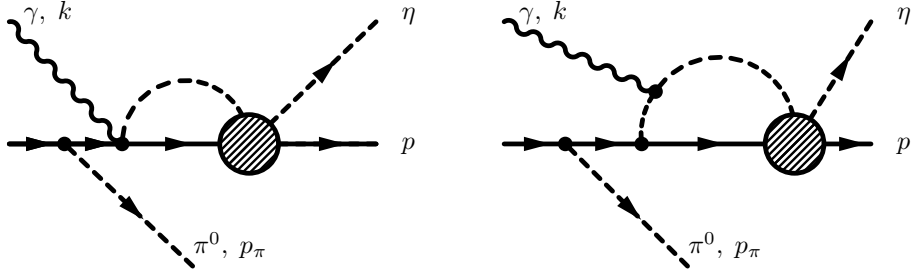}
\caption{Kroll-Ruderman term and meson pole term as sub-processes in the
$\pi^0\eta$ production.}
\label{fig:krorumepopio}
\end{figure}

Next, we take into account diagrams which involve the $\gamma N\to\eta N$ amplitude which has been discussed in Sec. \ref{sec_onemeson}, and which are shown in
 Fig. \ref{fig:krorumepopio}.
 With Eqs. (\ref{krmp}) and (\ref{alloneprod}) the amplitude for
the diagrams in Fig. \ref{fig:krorumepopio} is given by
\be (-i T_{\gamma p\to \pi^0\eta p})= \frac{D+F}{2 f_\pi}\;\frac{M}{E_N({\bf k+p_\pi})}\;\frac{i}{E_N({\bf k})-p_\pi^0-E_N({\bf
k+p_\pi})}\;(-i T_{\gamma p\to\eta p}(z))\;(-{\vec 
\sigma}\cdot {\bf
p_\pi})
\label{krmpres}
\ee
where $z$ takes the values given by Eq. (\ref{z1}) or (\ref{z2}).

\subsubsection{Intermediate pion emission}
\label{sec:intermediate}
\begin{figure}
\includegraphics[width=12cm]{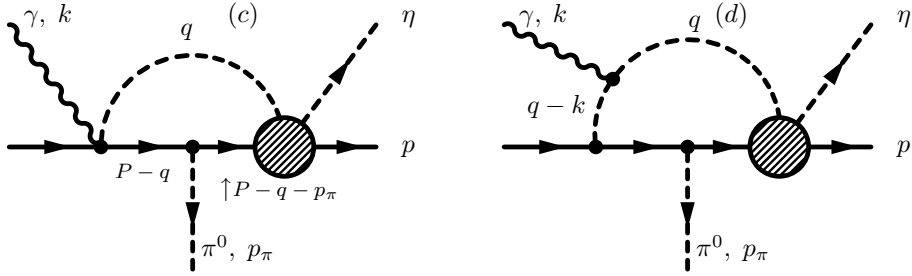}
\caption{Pion emission from inside the first meson-baryon loop. 
Diagram (d) is required
by gauge invariance.}
\label{fig:intermediate}
\end{figure}

In addition to the diagrams considered above, there are additional diagrams in which the
$\pi^0$ is produced inside the first meson-baryon loop as displayed in Fig.
\ref{fig:intermediate}.
The amplitude for the channel $i$ for the sum of diagram (c) and (d) 
is given by
\be t^i_{(c)+(d)}(\sqrt{s})&=&-\frac{e}{2f_\pi f_i} ({\vec \sigma}\cdot {\bf
p_\pi})({\vec \sigma} {\vec \epsilon})\: [a_i\: (D+F) + b_i \:(D-F)]
[a'_i \:(D+F) + b'_i \:(D-F)] \nonumber\\ &&\times\; T^{(i 3)}(z) 
\int\frac{d^3{\bf
q}}{(2\pi)^3}\;\frac{1}{2\omega}\;\frac{M}{E_M(q)}\;\frac{1}{\sqrt{s}-\omega-E_M(q)+
i\epsilon}\frac{M'}{E_{M'}({\bf
q+p}_\pi)}\;\non
&&\times\;\frac{1}{\sqrt{s}-\omega-p_\pi^0-E_{M'}({\bf
q+p}_\pi)+i\epsilon}\left(1-\frac{{\vec q}_{\rm on}^{\;2}}{3\:q_{\rm 
on}^0 \:k^0}\right)
F_\pi(q-k)
\label{intermediate}
\ee
where the index $i$ stands for our standard ordering of the channels in Eq.
(\ref{channels}) and the only non-zero values of the $a_i, a'_i, b_i, 
b'_i$ are: $a_2 =
-\frac{1}{\sqrt{2}}, a'_2 = -1, a_4=b_4=\frac{1}{\sqrt{6}}, a'_4 =
\sqrt{\frac{2}{3}}, b'_4 = -\frac{1}{\sqrt{6}}, a_5 = b_5 = \frac{1}{\sqrt{6}},
b'_5 = -\frac{1}{\sqrt{2}}$. Note that channel two has the external 
$\pi^0$ coupled to
$n$,
$n$ to the left and right in the diagram, channel four has the 
$\pi^0$ coupled to the
$\Lambda$,
$\Sigma$ to the left and right, and channel five has the $\pi^0$ 
coupled to the $\Sigma$,
$\Lambda$ to the left and right.  In the equation the variable $P-q$ 
refers to the baryon
on the left (M) of the emitted $\pi^0$ and the variable $P - q - 
p_\pi$ to the right
($M'$) of the emitted
$\pi^0$ as shown in Fig. \ref{fig:intermediate}.
The contribution of the terms in Fig. \ref{fig:intermediate} is 
therefore given by the
sum of Eq. (\ref{intermediate}) for the three non-vanishing channels.

In Eq. (\ref{intermediate}) we introduce the ordinary meson-baryon 
form factor $F_\pi$
of monopole type with $\Lambda$= 1.25 GeV as used in the two pion 
photoproduction
\cite{Nacher:2000eq}.  It appears naturally in the meson pole term of Fig.
\ref{fig:intermediate} and, as done in Ref. \cite{Nacher:2000eq}, it 
is also included in
Fig. \ref{fig:intermediate}(c) (Kroll-Ruderman term) for reasons of 
gauge invariance.
This form factor does not change much the results and it is 
approximated by taking the
$q^0$ variable on shell and making an angle average of the $\vec q$ 
momentum.  This is
done to avoid fictitious poles in the $q^0$ integrations.

The meson pole term (d) in Fig. \ref{fig:intermediate} is small and 
an approximation
can be made for the intermediate pion at $q-k$ which is far 
off-shell. This concerns
terms with mixed scalar products of the form ${\bf k\cdot q}$ that give only a small
contribution when integrating over ${\bf q}$ in Eq. 
(\ref{intermediate}). Additionally,
we have set in this term $q\equiv q_{\rm on}$, the on-shell momentum 
of the other meson
at $q$. This is, considering the kinematics, a good approximation.

One can also have the pion emission from the final proton.  However, 
this would imply
having the $\pi N\to \eta N$ amplitude away from the $N^*(1535)$ 
resonance at a value
$M_I = \sqrt{s}$ where the $\pi N \to \eta N$ amplitude would only 
provide a background
term above the $N^* (1535)$ resonance.  Once again the set of 
diagrams considered leads
to small cross sections compared to the dominant terms to be 
considered later in the
paper so further refinements are unnecessary.

\subsection{Baryonic resonances in $\eta\pi^0$ production}
\label{sec:resonances}
\begin{figure}
\includegraphics[width=12cm]{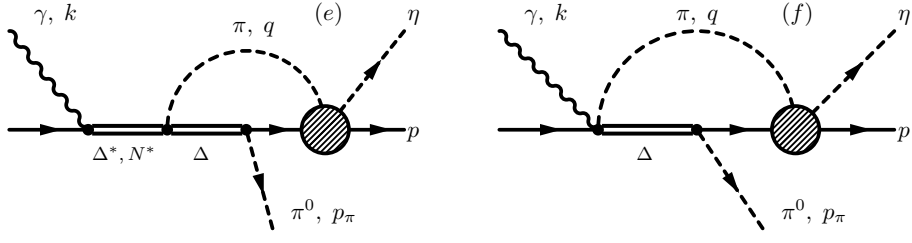}
\caption{Terms with $\Delta^*(1700)$, $N^*(1520)$, and 
$\Delta(1232)$. The diagram on the
right is the $\Delta$-Kroll-Ruderman term. The latter diagram implies 
also a meson pole
contribution, required by gauge invariance, which is not separately 
plotted but is
included in the calculation.}
\label{fig:deltas}
\end{figure}

In the present study, the $\eta\pi^0$ production is described as a
two-step process: The first step consists in the photoproduction of 
two mesons and a
baryon; the second step describes the subsequent transitions of 
meson-baryon $\to \eta
p$ via the dynamically generated $N^*(1535)$ resonance. In 
particular, the first stage
contains two pion photoproduction. For this part it is known that 
baryonic resonances
such as $\Delta$'s and $N^*$'s can play an important role
\cite{Nacher:2000eq,GomezTejedor:1993bq,GomezTejedor:1995pe,Fix:2005if}. 
For this reason
we include the relevant mechanisms from Ref. \cite{Nacher:2000eq} 
adapted to the present
context. Fig. \ref{fig:deltas} shows the processes that are taken into account.
The $s$-wave character of the $N^*(1535)$ (gray blob in Fig.
\ref{fig:deltas}) discards all those processes from  Ref. 
\cite{Nacher:2000eq} where both
pions couple in $p$-wave to the baryons, because the $\pi N$ loop 
function involving odd
powers of $\vec q$ in the integral is zero in the heavy baryon limit. 
For the remaining
processes, some contributions to the
$\pi^+\pi^0$ and $\pi^0\pi^0$ cross sections are small as, e.g., from the Roper
resonance. Finally, one is left with the $\Delta^*(1700)\Delta$,
$N^*(1520)\Delta$, and $\Delta$-Kroll-Ruderman terms from Fig. 
\ref{fig:deltas}. The
latter implies also a pole term which is required by gauge invariance 
in the same way as
in Fig. \ref{fig:intermediate} (d). Since many resonances appear in this section we refer 
the reader to the notation used in the Introduction.

The amplitudes for diagrams (e) and (f) in Fig. \ref{fig:deltas} are given by
\be
T_{\gamma p\to \pi^0\eta
p}=\sum_{i=1,2}T^{(i,3)}(z)\frac{1}{(2\pi)^2}\int\limits_0^{\Lambda} 
dq\int\limits_{-1}^1
dx\;t_\Delta^i\;\frac{q^2}{2\omega}
\frac{M}{E}\frac{1}{\sqrt{s}-\omega-p_\pi^0-E+i\epsilon}\;\frac{1}{\sqrt{s_\Delta}-M_\Delta+i\;\frac{\Gamma(\sqrt{s_\Delta})}{2}}
\label{full_tdeltas}
\ee
with the sum only over the first two channels according to Eq. 
(\ref{channels}) and
$\Lambda=1400$ MeV as in Sec. \ref{sec_onemeson}. The meson energy $\omega$, 
baryon energy $E$ and energy of the
$\Delta(1232)$ read
\be
\omega^2&=&m^2+q^2,\non
E^2&=&M^2+q^2+p_\pi^2+2q p_\pi x,\non
s_\Delta&=&\left(\sqrt{s}-\omega\right)^2-q^2
\ee
with meson mass $m$, baryon mass $M$, and $p_\pi=|{\bf p}_\pi|$ with
$p_\pi^0=\sqrt{p_\pi^2+m_\pi^2}$. The argument $z$ is given by Eq. 
(\ref{z1}) or
(\ref{z2}).

Using the notation from Ref. \cite{Nacher:2000eq} the amplitudes 
$t_\Delta^i$, which can
depend on the loop momentum, are given by:
\be
t^1_\Delta&=& t_{\gamma p\to\pi^+\pi^0 n}^{\Delta^*(1700)} + 
t_{\gamma p\to\pi^+\pi^0
n}^{N^*(1520)} + t_{\gamma p\to\pi^+\pi^0 n}^{\Delta-{\rm KR}}\;,\\
t^2_\Delta&=& t_{\gamma p\to\pi^0\pi^0
n}^{\Delta^*(1700)} +t_{\gamma p\to\pi^0\pi^0
n}^{N^*(1520)} +t_{\gamma p\to\pi^0\pi^0 p}^{\Delta-{\rm KR}}\ee
where
\be
  t_{\gamma p\to\pi^+\pi^0
n}^{\Delta^*(1700)}&=&-i\;\frac{2}{\sqrt{3}}\;\frac{f^*_{\Delta
N\pi}}{m_\pi}\;{\vec S}\cdot {\bf p}_\pi
\left({\tilde f}_{\Delta^*\Delta\pi}+\frac{1}{3}\frac{{\tilde
g}_{\Delta^*\Delta\pi}}{m_\pi^2}\;{\vec 
q}^{\;2}\right)G_{\Delta^*}(\sqrt{s}\:)\non
&&\left[g_1'\;\frac{{\vec S}^\dagger\cdot {\bf 
k}}{2M}\;(\vec{\sigma}\times {\bf
k})\vec{\epsilon}-i{\vec S}^\dagger\cdot{\vec 
\epsilon}\left(g_1'(k^0+\frac{{\bf
k}^2}{2M})+g_2'\sqrt{s}\:k^0\right)\right],
\label{deltastarampl}
\\
t_{\gamma p\to\pi^+\pi^0 
n}^{N^*(1520)}&=&-i\;\frac{\sqrt{2}}{3}\;\frac{f^*_{\Delta
N\pi}}{m_\pi}\;{\vec S}\cdot {\bf p}_\pi
\left({\tilde f}_{N^{*'}\Delta\pi}+\frac{1}{3}\frac{{\tilde
g}_{N^{*'}\Delta\pi}}{m_\pi^2}\;{\vec q}^{\;2}\right)G_{N^{*'}}(\sqrt{s}\:)\non
&&\left[g_1\;\frac{{\vec S}^\dagger\cdot {\bf k}}{2M}\;(\vec{\sigma}\times {\bf
k})\vec{\epsilon}-i{\vec S}^\dagger\cdot{\vec \epsilon}\left(g_1(k^0+\frac{{\bf
k}^2}{2M})+g_2 \sqrt{s}\:k^0\right)\right],
\label{nstarcontr}
\\
t_{\gamma p\to\pi^+\pi^0 n}^{\Delta-{\rm
KR}}&=&\frac{e\sqrt{2}}{9}\;\left(\frac{f^*_{\Delta N\pi}}{m_\pi}\right)^2
\left(2{\bf p}_\pi-i({\vec\sigma}\times{\bf p_\pi})\right)\cdot{\vec\epsilon}
\;F_\pi(q_{on}-k)\left(1-\frac{1}{3}\frac{{\vec q}_{\rm on}^{\;2}}{q_{\rm on}^0
k^0}\right),
\label{delkroru}
\\  t_{\gamma p\to\pi^0\pi^0
n}^{\Delta^*(1700)}&=&\frac{1}{2\sqrt{2}}\;t_{\gamma p\to\pi^+\pi^0
n}^{\Delta^*(1700)},
\label{deltastarampl2}
\\
t_{\gamma p\to\pi^0\pi^0
n}^{N^*(1520)}&=&\sqrt{2}t_{\gamma p\to\pi^+\pi^0 n}^{N^*(1520)},
\label{nstarcontr2}
\\
t_{\gamma p\to\pi^0\pi^0 p}^{\Delta-{\rm KR}}&=&0.
\label{end_deltas}
\ee
We have already projected out the $s$-wave parts of the 
$\Delta^*(1700)\Delta\pi$ and
$N^*(1520)\Delta\pi$ transitions that come from the term ${\tilde
g}_{N^{*'}\Delta\pi}/m_\pi^2\;{\vec S}^\dagger\cdot{\vec 
p}_\pi\;{\vec S}\cdot{\vec
p}_\pi$, see Ref. \cite{Nacher:2000eq}. The vector ${\bf p}_\pi$ 
depends implicitly on
the invariant mass which will be specified later, Eqs. (\ref{boost}) 
or (\ref{boost2}).
The amplitudes in Eqs. (\ref{deltastarampl})-(\ref{end_deltas}) are 
formulated for real
photons, which is the case we are considering here. The meson pole 
diagram related to the
$\Delta$-Kroll-Ruderman term has been included in the last factor of 
Eq. (\ref{delkroru})
by making the same approximation as in Eq. (\ref{intermediate}) for 
the intermediate
off-shell pion. The pion form factor $F_\pi$ (see Ref. 
\cite{Nacher:2000eq}) has to be
inserted since the intermediate pion in the meson pole term is far 
off-shell.  For the
$\Delta^*$ propagator, 
\be G_{\Delta^*}(\sqrt{s}\:)=\frac{1}{\sqrt{s}-M_{\Delta^*}+i\;\frac{\Gamma
(\sqrt{s}\:)}{2}},
\label{deltastarprop}
\ee 
the (momentum-dependent) width according to its main decay 
channels has been taken
into account: For $\Delta^*\to\pi N$ in $d$-wave, $\Delta^*\to N\rho 
(N\pi\pi)$, and
$\Delta^*\to \Delta\pi (N\pi\pi)$ we obtain in a similar way as in Ref.
\cite{Nacher:2000eq}
\be
\Gamma_{\Delta^*\to N\pi}(\sqrt{s}\:)&=&\Gamma_{\Delta^*\to N\pi}
(M_{\Delta^\star})\;\frac{q_{\rm CM}(\sqrt{s}\:)^5}{q_{\rm 
CM}(M_{\Delta^*})^5},\non
\Gamma_{\Delta^*\to
N\rho[\pi\pi]}(\sqrt{s}\:)&=&\frac{M_N}{6(2\pi)^3}\frac{m_{\Delta^*}}{\sqrt{s}}g_\rho^2
f_{\rho}^2\int d\omega_1\;d\omega_2\;|D_{\rho}(q_1+q_2)|^2 ({\bf 
q}_1-{\bf q}_2)^2
\Theta(1-|A|),\non 
A&=&\frac{(\sqrt{s}-\omega_1-\omega_2)^2-M_N^2-{\bf q}_1^2-{\bf
q}_2^2}{2|{\bf q}_1||{\bf q}_2|},\non
\Gamma_{\Delta^*\to\Delta\pi[N\pi\pi]}&=&\frac{15}{16\pi^2}\int
dM_I\;\frac{M_I\;k(M_I)}{4\pi \sqrt{s}\:}\;\frac{\Gamma_{\Delta\to
N\pi}(M_I)\left(|A_s|^2+|A_d|^2\right)}{(M_I-M_\Delta)^2+\left(\frac{\Gamma_{\Delta\to
N\pi}(M_I)}{2}\right)^2} \;\Theta(\sqrt{s}-M_I-m_\pi)\; .\non
\ee
Here, $q_{\rm CM}(\sqrt{s}\:)$ is the CM momentum of the pion and the 
nucleon and
$\Gamma_{\Delta^*\to N\pi} (M_{\Delta^\star})$ is determined through 
the branching ratio
into that channel. For the decay into $N \rho$, $g_\rho=2.6$ is the 
$\Delta^* N\rho$
coupling, also determined through the branching ratio. Furthermore, 
$f_\rho=6.14$ is the
$\rho\pi\pi$ coupling, $q_i=(\omega_i,{\bf q}_i)$, $i=1,2$ the 
four-momentum of the
outgoing pions, and $D_{\rho}$ the $\rho$ propagator incorporating the $\rho$ width. For 
the decay into
$\Delta\pi$, the finite width of the $\Delta$, $\Gamma_{\Delta\to 
N\pi}$, has been taken
into account by performing the convolution. For the partial 
amplitudes $A_s$ and $A_d$ of
the $\Delta^*$ decay into $\Delta\pi$ in $s$ and $d$-wave, see Ref.
\cite{Nacher:2000eq}.  The $N^*(1520)$ propagator is dressed in a 
similar way with the
analytic expressions given in Ref. \cite{Nacher:2000eq}.

\subsection{$SU(3)$ couplings of the $\Delta^*(1700)$ and background terms}
\label{sec:su3delstar} 
\begin{figure}
\includegraphics[width=12cm]{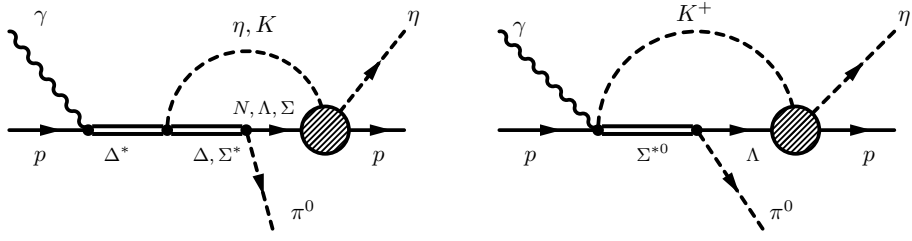}
\caption{Left side: Coupling of the dynamically generated 
$\Delta^*(1700)$ to $K
\Sigma^*$ and $\eta \Delta$. The loop is given by $\eta\Delta^+p$,
$K^+\Sigma^{*0}\Lambda$, or $K^0\Sigma^{*+}\Sigma^+$. Right side: $\Sigma^*$
Kroll-Ruderman term. }
\label{fig:sigstardelstar}
\end{figure}

In Ref. \cite{Sarkar:2004jh} the rescattering 
of the $0^-$ meson
octet with the $3/2^+$ baryon decuplet leads to a set of dynamically generated
resonances, one of which has been identified with the 
$\Delta^*(1700)$. The advantage of
such a microscopic model is that couplings of the resonance to decay 
channels are
predicted which have not yet been
determined experimentally. In particular, the analytic continuation 
of the amplitude to
the complex plane  provides at the pole position the isospin $3/2$ 
couplings of the
resonance to
$\eta\Delta$ and
$K\Sigma^*$. Identifying the pole with the $\Delta^*(1700)$ we can 
incorporate the model
from Ref.
\cite{Sarkar:2004jh} in the present study in the diagrammatic way as 
indicated on the
left side of Fig. \ref{fig:sigstardelstar}, with the $\gamma 
p\Delta^*$ coupling from
Ref. \cite{Nacher:2000eq}.
This procedure can be regarded as a first step towards the 
incorporation of dynamically
generated $3/2^-$ resonances in the two meson photoproduction. In 
further studies, the
initial $\gamma p\to\Delta^*$ process could be  included in the 
microscopic model of
Ref. \cite{Sarkar:2004jh} in a similar way as was done here for the 
$\gamma p N^*(1535)$
coupling in Sec. \ref{sec_onemeson}.  However, phenomenologically, 
the procedure
followed here is reliable.

The $\Delta^*(1700)\Delta\eta$ and $\Delta^*(1700)\Sigma^* K$ couplings from Ref.
\cite{Sarkar:2004jh} are  given up to a global sign by $g_\eta=1.7-i 
1.4$ and $g_K=3.3+i
0.7$, respectively.  However, in Ref. \cite{Sarkar:2004jh} the 
coupling to $\Delta\pi$
is also given, $g_\Delta=0.5+i\;0.8$.  The sign of the real part and 
the order of magnitude agree with the
empirical analysis of the $\Delta^*(1700) \to \pi\Delta$ decay that 
we are using thus far \cite{Nacher:2000eq};
hence, we take for $g_\eta$ and $g_K$ the values quoted above.   We 
note
that the cross section is almost independent of the global sign, 
whereas there are some minor
differences in the invariant mass spectra.

Having included the $\Sigma^*$ in the $\Delta^*$ decay it is 
straightforward to consider
also the corresponding $\Sigma^*$-Kroll-Ruderman term  given on the 
right side of Fig.
\ref{fig:sigstardelstar}. This term, together with the other ones 
from this section,
allows for an extension of the model to higher energies, where the intermediate
$\Delta(1232)$ from the processes of Sec. \ref{sec:resonances} is 
off-shell but the
$\Sigma^*(1385)$ is on-shell. 

For the baryon decuplet baryon octet meson octet vertices, and the corresponding Kroll-Ruderman vertex,
we take the effective Lagrangian from Ref. \cite{Butler:1992pn},
\be
{\cal L}={\cal C}\left(\overline{T}_\mu A^{\mu}B+{\overline B}A_\mu T^\mu\right)
\ee
with the same phase phase conventions for the states of the decuplet as taken there, which is the same 
one taken in Ref. \cite{Sarkar:2004jh}. This allows us to relate all the couplings to the one of $\Delta\pi N$. 
Up to a different phase, these factors agree with those used in Ref. \cite{Oset:2000eg}.

The corresponding amplitudes for the diagrams in Fig. \ref{fig:sigstardelstar} read now:
\be
t_{\eta\Delta^+ p}^{(3)}&=&-\sqrt{\frac{2}{3}}\;g_\eta\;\frac{\;f^*_{\Delta
N\pi}}{m_\pi}\; G_{\Delta^*}(\sqrt{s})\;{\vec S}\cdot {\bf
p}_\pi\left[-ig_1'\;\frac{{\vec S}^\dagger\cdot {\bf 
k}}{2M}\;(\vec{\sigma}\times {\bf
k})\vec{\epsilon}-{\vec S}^\dagger\cdot{\vec
\epsilon}\left(g_1'(k^0+\frac{{\bf k}^2}{2M})+g_2'\sqrt{s}\;k^0\right)\right],
\label{deltastarbgresc}
\\
t_{K^+\Sigma^{*0}\Lambda}^{(5)}&=&1.15\sqrt{\frac{24}{25}}\;g_K\;\frac{D+F}{2f_\pi}\;G_{\Delta^*}(\sqrt{s})\vec{S}\cdot
{\bf p}_\pi\left[-ig_1'\frac{\vec{S}^\dagger\cdot {\bf 
k}}{2M}(\vec{\sigma}\times {\bf
k})\cdot\vec{\epsilon}-\vec{S}^\dagger\cdot\vec{\epsilon}\left(g_1'(k^0+\frac{{\bf
k}^2}{2M})+g_2'\sqrt{s}\;k^0\right)\right],
\label{delstarksig1}
\\ t_{K^0\Sigma^{*+}\Sigma^+}^{(6)}&=&\frac{2}{5}\frac{D+F}{2f_\pi}\;g_K\;
G_{\Delta^*}(\sqrt{s})\vec{S}\cdot {\bf 
p}_\pi\left[-ig_1'\frac{\vec{S}^\dagger\cdot {\bf
k}}{2M}(\vec{\sigma}\times {\bf
k})\cdot\vec{\epsilon}-\vec{S}^\dagger\cdot\vec{\epsilon}\left(g_1'(k^0+\frac{{\bf
k}^2}{2M})+g_2'\sqrt{s}\;k^0\right)\right],
\label{delstarksig2}
\\ t_{\Sigma^*{\rm
-KR}}^{(5)}&=&-\;1.15\;e\;\frac{4\sqrt{3}}{25}\left(\frac{D+F}{2f_\pi}\right)^2\left(2{\bf
p}_\pi-i(\vec{\sigma}\times {\bf p}_\pi)\right)\cdot\vec{\epsilon}
\label{new_sigstars}
\ee  
with $g_\eta$ and $g_K$ given in Ref. \cite{Sarkar:2004jh}. In 
order to obtain the
full amplitudes, $T_{\gamma p\to \pi^0\eta p}$, these $t$'s have to be 
inserted as
$t_\Delta^i$ Eq. (\ref{full_tdeltas}) but the sum over index $i$ goes 
now from three to
six. The lower index for the amplitudes in Eqs.
(\ref{deltastarbgresc}) - (\ref{new_sigstars}) indicates the 
particles in the loop to
be considered in the evaluation of Eq. (\ref{full_tdeltas}). The 
upper index indicates
the channel number $i$ and therefore which $T^{(i3)}$ has to be chosen in Eq.
(\ref{full_tdeltas}). For the amplitudes from Eq. (\ref{delstarksig1}),
(\ref{delstarksig2}), and (\ref{new_sigstars}), the
$\Delta(1232)$ propagator in Eq. (\ref{full_tdeltas}) has to be 
replaced with the
$\Sigma^*(1385)$ one. The latter is defined in the same way as the
$\Delta(1232)$ propagator and we take a momentum-dependent width with
$\Gamma_{\rm rest}=36$ MeV assuming the dominant $p$-wave decay of 
the $\Sigma^*$ into
$p \Lambda$.
The numerical factor of 1.15 appearing in Eqs. (\ref{delstarksig1}) 
and (\ref{new_sigstars})
is a phenomenological correction factor from the SU(3) $\Sigma^* \pi 
\Lambda$ coupling
in order to provide the empirical $\Sigma^* \to \pi\Lambda$ partial 
decay width.

\begin{figure}
\includegraphics[width=6cm]{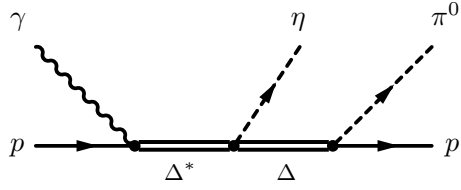}
\caption{Tree level process from the decay of the $\Delta^*(1700)$ 
to $\eta\Delta(1232)$.}
\label{fig:background}
\end{figure}

For the $\pi^0 \eta p$ production, the $\Delta^*\Delta\eta$ coupling 
together with the
subsequent $\Delta\to\pi^0 p$ decay provides also a term 
at tree level as
shown in Fig. \ref{fig:background}.
The contribution for this reaction is simply given by
\be T_{\gamma p\to\pi^0\eta p}^{\rm BG}&=&t_{\eta\Delta^+ 
p}^{(3)}\;G_\Delta (z')
\label{deltastarbg}
\ee from Eq. (\ref{deltastarbgresc}). The invariant argument $z'$ for 
this amplitude
amplitude differs from $z$ of the former processes,
\be z'=\left(s+m_\eta^2-2\sqrt{s}\;p_\eta^0\right)^{1/2},\quad z'=M_I
\label{z3}
\ee with $p_\eta^0=({\bf p}_\eta^2+m_\eta^2)^{1/2}$ depending on 
whether the amplitude is
parametrized in terms of $M_I(\eta p)$ or $M_I(\pi^0p)$, 
respectively. We have explicitly
tested that recoil corrections for the $\Delta(1232)$ decay ${\vec 
S}\cdot {\bf p}_\pi$
in Eq. (\ref{deltastarbg}), in the way they are applied in Ref. 
\cite{Nacher:2000eq}, are
negligible.

\section{Results}
\label{sec:results_1} In this section, invariant mass spectra $M_I(\eta p)$ and
$M_I(\pi^0 p)$ for the reaction $\gamma p\to \pi^0\eta p$ are 
predicted, together with the
total cross section for this reaction. The corresponding observables 
for the $\pi^0K^0\Sigma^+$
final state are also given. These observables can be directly 
compared to ongoing experiments at
the ELSA facility\cite{Metag}.

We evaluate the phase space integrals for the invariant mass 
distribution of $\eta p$ in
the $\eta p$ CM system,
\be
\frac{d\sigma}{dM_I(\eta p)}=\frac{1}{4(2\pi)^5}\frac{M_p
M_i}{s-M_p^2}\;\frac{\tilde{p}_\eta
p_\pi}{\sqrt{s}}\int\limits_0^{2\pi}d\phi_\pi\int\limits_{-1}^1d\cos\theta_\pi\int\limits_0^{2\pi}d\tilde{\phi}\int\limits_{-1}^1d\cos\tilde{\theta}\;
\overline{\sum}\sum |T_{\gamma p\to\pi^0\eta p}|^2\non
\label{phase_space}
\ee
with $\tilde{p}_\eta$ the modulus of the momentum ${\vec {\tilde 
p}}_\eta$ of the $\eta$
in the $\eta p$ rest frame ${\tilde p}_\eta=\lambda^{1/2}(M_I^2, 
m_\eta^2,M_p^2)/(2M_I)$
in terms of the ordinary K\"allen function where the direction of
${\vec {\tilde p}}_\eta$ is given by $\tilde{\phi}$ and 
$\tilde{\theta}$. This vector is
connected to ${\vec p_\eta}$ in the $\gamma p$ rest frame by the boost
\be {\vec 
p}_\eta=\left[\left(\frac{\sqrt{s}-\omega_\pi}{M_I}-1\right)\left(-\frac{{\vec
{\tilde p}}_\eta{\vec p}_\pi}{{\vec p}_\pi^{\;2}}\right)+\frac{{\tilde
p}_\eta^0}{M_I}\right]\left(-{\vec p}_\pi\right)+{\vec {\tilde p}}_\eta
\label{boost}
\ee
where ${\tilde p}_\eta^0=\sqrt{{\vec {\tilde p}}_\eta^{\;2}+m_\eta^2}$ and the
$\pi^0$ three momentum in the $\gamma p$ CM frame is given by the modulus
$p_\pi=\lambda^{1/2}(s,M_I^2, m_\pi^2)/(2\sqrt{s})$ and the two angles
$\phi_\pi,\theta_\pi$. Furthermore, $\omega_\pi$ is the pion energy 
in the $\gamma p$ CM
frame. In Eq. (\ref{phase_space}),
$M_p, M_i$ are proton mass and mass of the final baryon, in the 
present case also a
proton ($i=3$ with the channel ordering from Eq. (\ref{channels})).  Eqs.
(\ref{phase_space}) and (\ref{boost}) are a generalization of the corresponding
expression in Ref. \cite{Nacher:1998mi} as in the present case the 
amplitude depends explicitly on the angles
of the particles.

\begin{figure}
\includegraphics[width=11cm]{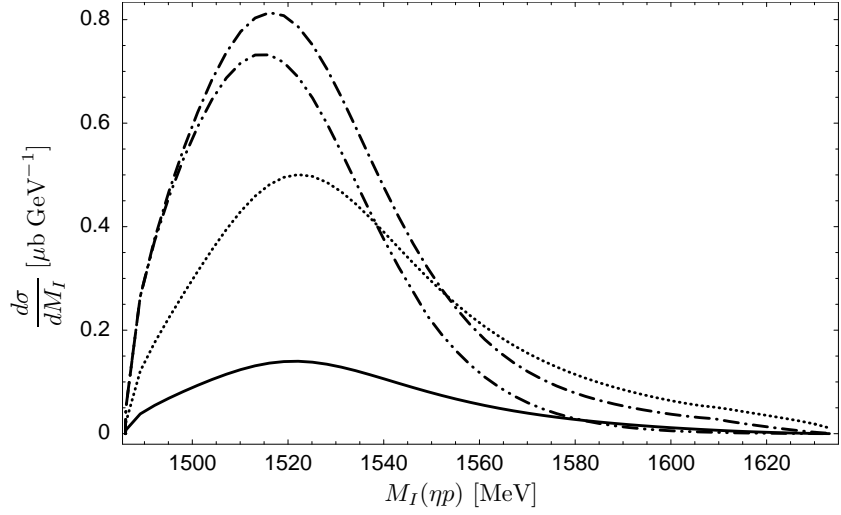}
\caption{Invariant mass at $E_\gamma=1.2$ GeV. Dotted
line:  Contact interaction from Fig. \ref{fig:rescattering_diagrams} 
including the anomalous magnetic
moment. Dashed dotted line:  Meson pole plus Kroll-Ruderman term from 
Fig.  \ref{fig:krorumepopio}.
Double dashed dotted line:
$\Delta^*(1700)K\Sigma^*$ transitions from Fig. 
\ref{fig:sigstardelstar}, see Eqs. (\ref{delstarksig1}), 
(\ref{delstarksig2}).
Solid line:  Intermediate pion emission from Fig. \ref{fig:intermediate}.}
\label{fig:dsdmsmall}
\end{figure}

\begin{figure}
\includegraphics[width=11cm]{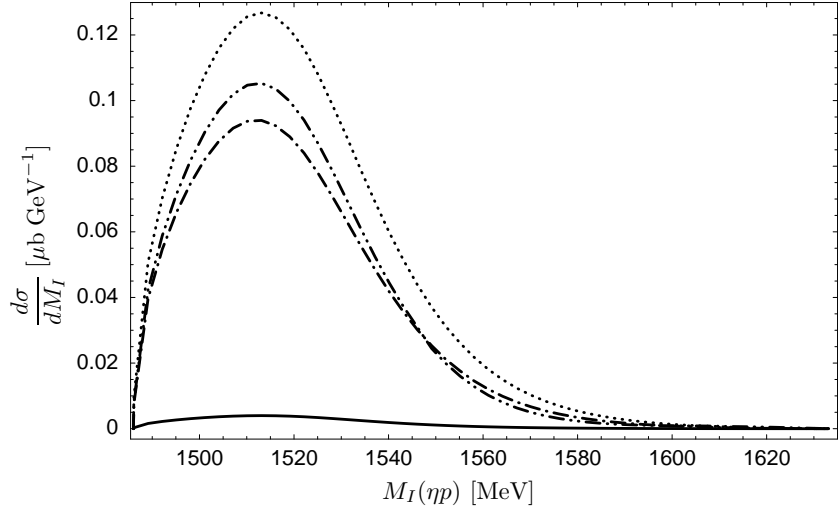}
\caption{Invariant mass at $E_\gamma=1.2$ GeV. Processes with 
explicit resonances.  Dotted line:
$\Delta^*(1700)\pi\Delta$ contribution from Fig. \ref{fig:deltas} 
(e), see Eqs. (\ref{deltastarampl}) and
(\ref{deltastarampl2}). Solid line: Contribution from 
$N^*(1520)\pi\Delta$ in Fig. \ref{fig:deltas} (e), see Eqs.
(\ref{nstarcontr}) and (\ref{nstarcontr2}). Dashed dotted line:
$\Delta$-Kroll-Ruderman term from Fig. \ref{fig:deltas} (f), see Eq. 
(\ref{delkroru}). Double dashed dotted line:
$\Sigma^*$-Kroll-Ruderman term from Fig. \ref{fig:sigstardelstar}, 
see Eq. (\ref{new_sigstars}).}
\label{fig:dsdmmiddle}
\end{figure}

\begin{figure}
\includegraphics[width=11cm]{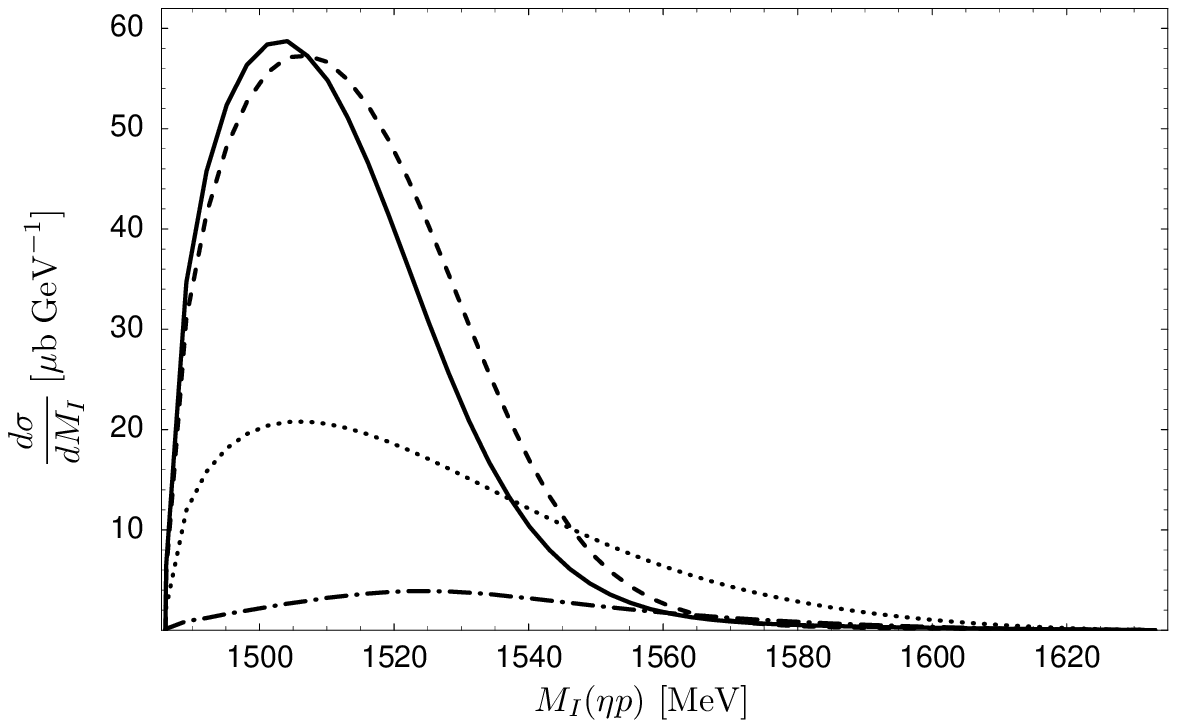}
\caption{Invariant mass at $E_\gamma=1.2$ GeV.  Dashed dotted line:
$\Delta^*(1700)\eta\Delta$ transition and $\eta p\to\eta p$ 
rescattering from Fig. \ref{fig:sigstardelstar}, see Eq. 
(\ref{deltastarbgresc}).
Dotted line:  Tree level process with $\Delta^*(1700)\eta \Delta$ 
transition but no rescattering,
Fig. \ref{fig:background}, see Eq. (\ref{deltastarbg}). Solid line: 
Coherent sum of all contributions, full model for the
$N^*(1535)$.  Dashed line: Coherent sum of all contributions, reduced
model for $N^*(1535)$ from Sec. \ref{sec_pinetan} (no vector 
particles in $t$-channel, no
$\pi\pi N$ channel).  }
\label{fig:dsdmlarge}
\end{figure}

The individual numerical contributions from the various processes from Sec.
\ref{sec_twofinal} are shown in Figs. \ref{fig:dsdmsmall}, 
\ref{fig:dsdmmiddle}, and
\ref{fig:dsdmlarge}. We have chosen here a lab energy for the photon 
of $E_\gamma=$1.2
GeV so that the allowed invariant mass range is wide enough to 
distinguish the $N^*(1535)$
from pure phase space. On the other hand, this energy is low enough, 
so that unknown
contributions from heavier resonances than the $\Delta^*(1700)$ 
should be small.
All contributions contain the resonant structure of the $N^*(1535)$ in the
final state interaction, except the background term from Eq. 
(\ref{deltastarbg}).
Although the shape of this contribution is similar to the resonant 
part, this is a
combined effect of phase space and the intermediate $\Delta(1232)$ 
that becomes less
off-shell at lower invariant masses for $E_\gamma=1.2$ GeV.

The individual contributions shown in Figs. \ref{fig:dsdmsmall}, 
\ref{fig:dsdmmiddle},
and \ref{fig:dsdmlarge} are evaluated using the  full model for the 
$N^*(1535)$ from Sec.
\ref{sec_pinetan} with the coherent sum indicated in Fig. 
\ref{fig:dsdmlarge}, solid
line.  The coherent sum using the reduced model is displayed with the 
dashed line in
Fig. \ref{fig:dsdmlarge}.  We take the difference between the two curves as an
indication of the theoretical uncertainty as in the previous sections.

The first thing to note is that the peak position of the $N^*(1535)$ 
is lowered by some
20 MeV due to the interference of the dynamically generated resonance 
with the background
term from Fig. \ref{fig:background}. A width of 93 MeV for the 
$N^*(1535)$ has been extracted in Ref.
\cite{Inoue:2001ip}. In the invariant mass spectra the $N^*(1535)$ exhibits a
considerably smaller width.  This is for two reasons: First, the 
$N^*(1535)$ is situated
close to the $\eta p$ threshold and the phase space cuts the lower 
energy tail.  This is
clearly visible in Fig. \ref{fig:only_resonance}:
\begin{figure}
\includegraphics[width=16cm]{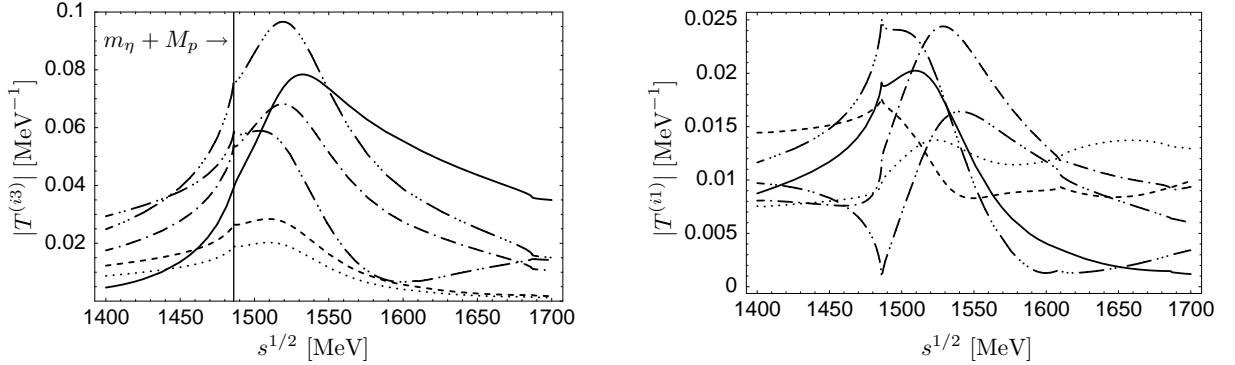}
\caption{Modulus of the amplitude (full model for $N^*(1535)$) for the reaction
$M_iB_i\to\eta p$ (left side) and $M_iB_i\to\pi^0 p$ (right side). As 
these amplitudes
appear squared in invariant mass spectra and total cross sections, 
they serve as a useful
tool to distinguish dominant processes for the $\pi^0\eta p$ final 
state. Initial
states:  Dotted lines: $\pi^0 p$, Dashed lines: $\pi^+ n$, Solid 
lines: $\eta p$, Dashed
dotted line: $K^+\Sigma^0$, Double dashed dotted lines: $K^+ 
\Lambda$, Triple dashed
dotted lines: $K^0 \Sigma^+$.}
\label{fig:only_resonance}
\end{figure} As the phase space factors in Eq. (\ref{phase_space}) are  smooth
functions around the $N^\star(1535)$ resonance, the shape of the 
curves in Figs.
\ref{fig:dsdmsmall}, \ref{fig:dsdmmiddle}, and
\ref{fig:dsdmlarge} reflects the $N^\star(1535)$ resonance seen through a
$|T|^2$ matrix involving the coupled channels.

\begin{figure}
\includegraphics[width=11cm]{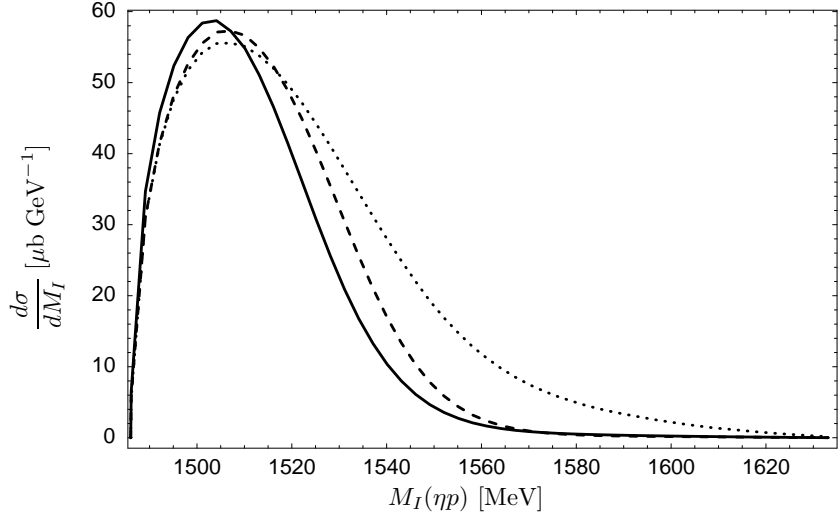}
\caption{Phenomenological potential for the $MB\to\eta p$ transition at 
$E_\gamma=1.2$ GeV.
With solid and dashed line, full and reduced model for the 
$N^*(1535)$ as in Fig.
\ref{fig:dsdmlarge}. Dotted line:  Phenomenological potential in the 
meson-baryon
$\to\eta p$ final state interaction for the diagrams from Figs.
\ref{fig:rescattering_diagrams}, \ref{fig:krorumepopio}, 
\ref{fig:intermediate},
\ref{fig:deltas}, and \ref{fig:sigstardelstar}.}
\label{fig:pheno_pot}
\end{figure}

The second reason for the narrow $N^\star(1535)$ is that at higher 
invariant mass the
amplitude for the resonance is suppressed by the initial 
photoproduction mechanism: A
closer inspection of the dominant resonant contributions as, e.g., from Eq.
(\ref{deltastarbgresc}) shows that the $\Delta(1232)$ propagator in 
Eq. (\ref{full_tdeltas}) of the first loop becomes more and
more off-shell at higher $\eta p$ invariant masses which leads to a 
suppression of the spectrum for
this kinematics. This effect is in fact so pronounced that the
shape of the invariant mass distribution hardly changes if the $M B \to \eta p$
theoretical transition amplitudes are replaced in the scattering 
diagrams by the
phenomenological ones given by Eq. (\ref{phenonepord}).  This is 
clearly seen in Fig.
\ref{fig:pheno_pot}.

The transitions $|T^{(i3)}|$ and $|T^{(i1)}|$ in Fig. 
\ref{fig:only_resonance} explain
the size of some of the contributions in Figs. \ref{fig:dsdmsmall}, 
\ref{fig:dsdmmiddle},
and \ref{fig:dsdmlarge} as they appear squared in invariant mass 
spectra and cross
section. E.g., the $\pi N\to\eta p$ transitions on the left side of Fig.
\ref{fig:only_resonance} are small which explains why the 
$\Delta$-Kroll-Ruderman term
and the $N^*(1525)\pi\Delta$, $\Delta^*(1700)\pi\Delta$ transitions from Eqs.
(\ref{deltastarampl})-(\ref{end_deltas}) contribute little, 
opposite to what was found
in the two-pion photoproduction \cite{Nacher:2000eq}. In 
contrast, the diagrams
using $SU(3)$ Lagrangians without explicit resonances from Figs.
\ref{fig:rescattering_diagrams}, \ref{fig:krorumepopio}, and 
\ref{fig:intermediate}
contain $K\Lambda$ and $K\Sigma$ channels
in the first loop so that the contributions are larger. The by far 
largest contributions
in the rescattering part (Fig. \ref{fig:dsdmlarge}) comes from the
$\Delta^*(1700)\to\eta\Delta$ decay with the subsequent unitarization of $\eta
p$. Indeed, the $\eta p\to\eta p$ scattering amplitude is very large as Fig.
\ref{fig:only_resonance} shows. Additionally, the loop for this 
reaction in Fig.
\ref{fig:sigstardelstar} contains a $\eta$ instead of a $\pi$, and 
the particles in the
loop can be simultaneously on-shell, whereas for the $\pi\Delta N$ 
loop at least one
particle is always further off-shell.

\begin{figure}
\includegraphics[width=15cm]{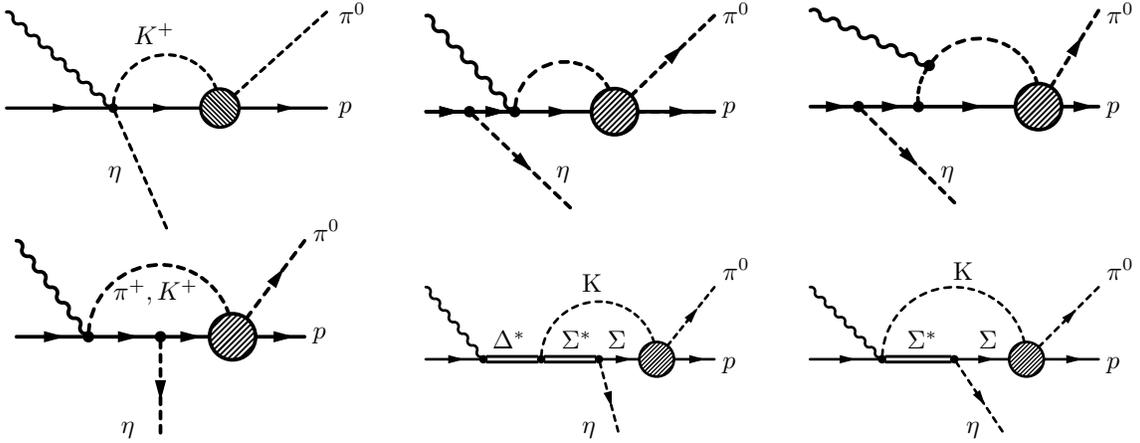}
\caption{Selection of diagrams with $\pi^0 p$ being the final state 
of the rescattering
instead of $\eta p$. These diagrams are suppressed.}
\label{fig:more_diagrams}
\end{figure}

The diagrams with $\Sigma^*(1385)$ in the first loop are relatively large (Fig.
\ref{fig:dsdmsmall}) due to the large $\Delta^*(1700)K\Sigma^*$ coupling and
the large $K\Sigma\to\eta p$ and $K\Lambda\to\eta p$ transitions from Fig. \ref{fig:only_resonance}. 
However, the 
$\Sigma^*(1385)$ is
off-shell at $E_\gamma=1.2$ GeV and the contribution can not become as big as 
the loop from the $\Delta^*(1700)\to\Delta\eta$ decay.
Therefore, diagrams with a $\Delta^*(1700)K\Sigma^*$ coupling become 
more important at 
higher energies.
From Fig. \ref{fig:only_resonance}, right side, we can also directly 
read off that
additional diagrams like those displayed in Fig. \ref{fig:more_diagrams}
which use $T^{(i1)}$ instead of $T^{(i3)}$ are small 
compared to their
counterparts from Sec. \ref{sec_twofinal}.

\subsection{Extension to higher energies}
\label{sec:extension_higher} 
\begin{figure}
\includegraphics[width=16cm]{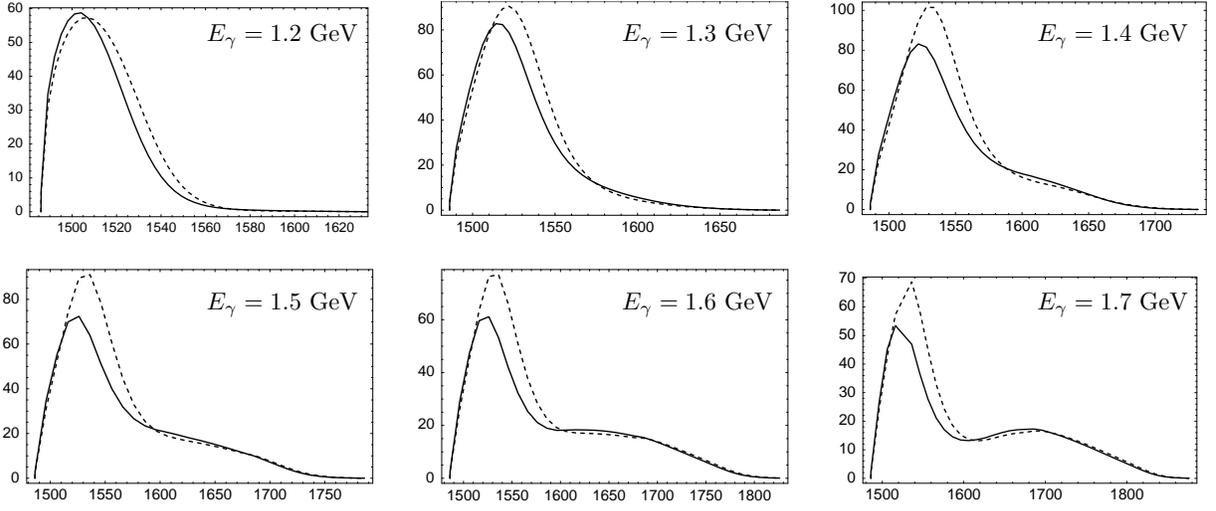}
\caption{Invariant mass spectrum $\frac{d\sigma}{dM_I(\eta 
p)}\;[\mu{\rm b\;GeV}^{-1}]$
as a function of $M_I(\eta p)$ [MeV] for various photon lab energies 
$E_\gamma$.  Solid and
dashed lines: Full and reduced model for the $N^*(1535)$, respectively.}
\label{fig:bigger_window}
\end{figure}

In Fig. \ref{fig:bigger_window}
 the results for the invariant mass distribution are 
shown for higher
values   of the incoming
$\gamma$ momentum, $q_{\rm lab}=1.2-1.7$ GeV.  The resonant shape of 
the $N^*(1535)$ is
not modified if a bigger photon energy is chosen, only the size 
decreases slightly as the
intermediate $\Delta^*(1700)$ of the dominant processes becomes off-shell.
At higher incident photon energies, the peak of the $N^*(1535)$ moves back
to  its original position around 1520-1540 MeV (see, e.g., Fig. 
\ref{fig:s11pinetan}) as the interference of the dynamically
generated $N^*(1535)$ with the tree level process from Fig. \ref{fig:background} becomes weaker.

A second maximum appears for $E_\gamma>\sim 1.5$ GeV and moves  to 
higher invariant
masses with increasing photon energy. This can be traced back  to be 
a reflection of the
$\Delta(1232)$ resonance in the tree level process from Fig. 
\ref{fig:background} which
is on-shell around  the position of the second peak.  When predicting 
this double hump
structure in the $\eta p$ invariant mass, one has to keep in mind 
that our model for the
dynamically generated
$N^*(1535)$ resonance underpredicts the width of this resonance (see, 
e.g., Figs.
\ref{fig:s11pinetan}, \ref{fig:csonemeson}). Furthermore, there are unknown
contributions from resonances heavier than the $\Delta^*(1700)$ about 
which little is
known and which can fill up the space in invariant mass between the 
two humps. As a
result, we expect a separation of the two maxima not at 
$E_\gamma=1.5$ GeV as Fig.
\ref{fig:bigger_window} suggests but at higher energies. 
Nevertheless, the tree level
process from Fig. \ref{fig:background} contributes so strongly to the 
coherent sum  that
the double hump structure should be qualitatively visible in experiment.

\begin{figure}
\includegraphics[width=8.5cm]{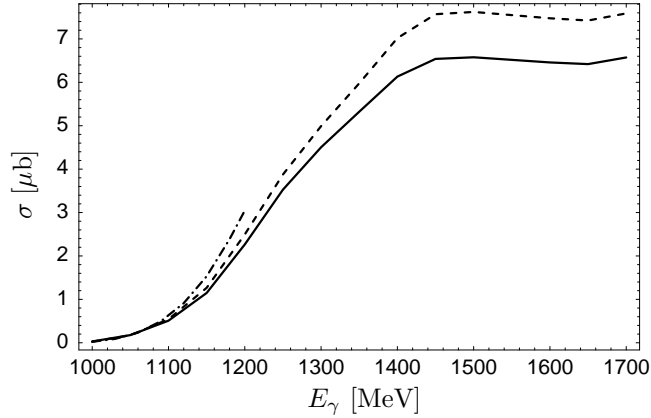}
\caption{Integrated cross section for the $\gamma p\to \pi^0 \eta p$ 
reaction. Solid
line: Full model for the
$N^*(1535)$.  Dashed line: Reduced model (see Sec. \ref{sec_pinetan}). 
Dashed dotted line: Phenomenological potential
for the $MB\to \eta p$ transition (only available up to $E_\gamma\sim 
1.2$ GeV).}
\label{fig:cross_section}
\end{figure}

In Fig. \ref{fig:cross_section}
  the integrated cross section is shown. There is a steep 
rise below
$E_\gamma=1.2$ GeV simply due to growing phase space. Above that, the 
cross section grows
slower and finally saturates. At high photon energies, the tree level 
process and the
dynamically generated $N^*(1535)$ are almost completely separated in 
invariant mass (see
Fig.
\ref{fig:bigger_window}) and we do not expect a further rise beyond 
1.7 GeV within our approach, as the
particles involved in the various processes become more and more off-shell.
However, the narrow $N^*(1535)$ width of our model, together with 
unknown contributions
from resonances heavier than the $\Delta^*(1700)$, lead to uncertainties at high photon energies which
are hard to control.

As we have already 
seen in Fig. \ref{fig:pheno_pot}, the use of the wider
phenomenological potential increases the cross section slightly (dashed dotted
line), but it is remarkable how insensitive the cross section is 
to the actual width of the $N^*(1535)$, regarding the large difference
in width between the results using the phenomenological potential or microscopic theory which
we have seen for the $\gamma p\to\eta p$ reaction in Fig. \ref{fig:csonemeson}.

\subsection{The $\pi^0 p$ invariant mass}
In the discussion of the last section we have
seen that the $\Delta(1232)$ plays a prominent role in the
$\pi^0\eta p$ photoproduction. For completeness, the invariant mass 
spectra for the
$\pi^0 p$ particle pair is given, which should show a signal of the
$\Delta(1232)$. The phase space integrals are evaluated in the $\pi^0 
p$ rest frame and
lead --- similar to the expression in Eq. (\ref{phase_space}) --- to 
the invariant mass
distribution for $M_I(\pi^0 p)$:
\be
\frac{d\sigma}{dM_I(\pi^0 p)}=\frac{1}{4(2\pi)^5}\frac{M_p
M_i}{s-M_p^2}\;\frac{\tilde{p}_\pi
p_\eta}{\sqrt{s}}\int\limits_0^{2\pi}d\phi_\eta\int\limits_{-1}^1d\cos\theta_\eta\int\limits_0^{2\pi}d\tilde{\phi}\int\limits_{-1}^1d\cos\tilde{\theta}\;
\overline{\sum}\sum |T_{\gamma p\to\eta\pi^0 p}|^2\non
\label{dsdmpidel}
\ee  
with $M_i=M_p$ and
with $\tilde{p}_\pi$ the modulus of the momentum  ${\vec {\tilde 
p}}_\pi$ of the
$\pi^0$ in the $\pi^0 p$ rest frame ${\tilde p}_\pi=\lambda^{1/2}(M_I^2,
m_\pi^2,M_p^2)/(2M_I)$ where the direction of ${\vec {\tilde p}}_\pi$ 
is given by
$\tilde{\phi}$ and $\tilde{\theta}$. This vector is connected to 
${\vec p_\pi}$ in the
$\gamma p$ rest frame by the boost
\be  {\vec 
p}_\pi=\left[\left(\frac{\sqrt{s}-\omega_\eta}{M_I}-1\right)\left(-\frac{{\vec
{\tilde p}}_\pi{\vec p}_\eta}{{\vec p}_\eta^{\;2}}\right)+\frac{{\tilde
p}_\pi^0}{M_I}\right]\left(-{\vec p}_\eta\right)+{\vec {\tilde p}}_\pi
\label{boost2}
\ee  where ${\tilde p}_\pi^0=\sqrt{{\vec {\tilde 
p}}_\pi^{\;2}+m_\pi^2}$ and the
$\eta$ three momentum in the $\gamma p$ CM frame is given by the modulus
$p_\eta=\lambda^{1/2}(s,M_I^2, m_\eta^2)/(2\sqrt{s})$ and the two angles
$\phi_\eta,\theta_\eta$.  Note that the invariant arguments for the 
solution of the
Bethe-Salpeter equation (\ref{bse}) have changed compared to the case when
the amplitude is expressed in terms of the $\eta p$ invariant mass,
see Eqs. (\ref{z1}), (\ref{z2}), and (\ref{z3}).

\begin{figure}
\includegraphics[width=17cm]{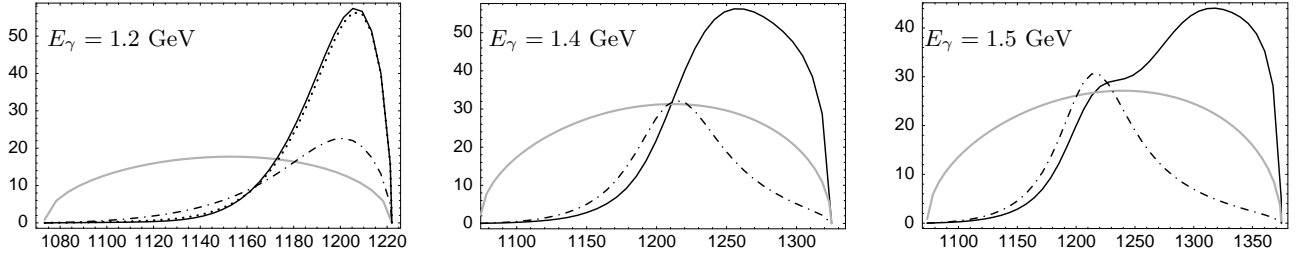}
\caption{Invariant mass spectrum $\frac{d\sigma}{dM_I(\pi^0 
p)}\;[\mu{\rm b\;GeV}^{-1}]$
as a function of $M_I(\pi^0p)$ [MeV] for various photon lab energies 
$E_\gamma$. Solid
lines: Full model for the $N^*(1535)$. Gray lines: Phase space only 
($T$=const).
Dashed dotted lines: Only tree level process from Fig. 
\ref{fig:background}, see Eq. (\ref{deltastarbg}).
Dotted line in plot for $E_\gamma=1.2$ GeV:  Effect when including
recoil corrections for the $\Delta\pi^0 p$ vertex for the
tree level diagram in Fig. \ref{fig:background}. }
\label{dsdmpip}
\end{figure}

The invariant mass distribution including all processes from this 
study is plotted in
Fig. \ref{dsdmpip}.
We have checked explicitly for the individual processes and for the
coherent sum of all processes that the integration over $M_I(\pi^0 p)$ in Eq.
(\ref{dsdmpidel}) leads to the same values for the cross section as 
when integrating over
the $\eta p$ invariant mass distribution from Eq. 
(\ref{phase_space}). In the plot for $E_\gamma=1.2$ GeV,
the dotted line indicates the negligible effect of recoil corrections 
for the tree
level process from Fig. \ref{fig:background} as described below Eq. (\ref{z3}).

In Fig. \ref{dsdmpip} we observe at $E_\gamma=1.2$ GeV a shift of 
strength towards higher
invariant masses  compared to the pure phase space (gray
line) obtained by setting $T=$ const in Eq. (\ref{dsdmpidel}). This 
is caused by the low
energy tail of the $\Delta(1232)$ from the tree level process from Fig.
\ref{fig:background}, indicated with the dashed-dotted line. Indeed, 
at higher photon
energies, the intermediate $\Delta(1232)$ in this process shows up as a 
shoulder at $E_\gamma=1.5$ GeV, and as a clear peak beyond.
Additionally, there is a shift of strength
towards higher invariant masses that results in a maximum which moves 
with energy, as it
becomes apparent at $E_\gamma=1.5$ GeV. This is a reflection of the 
$N^*(1535)$ resonance
that becomes on-shell around these invariant masses, in full analogy 
to the reflection of the
$\Delta(1232)$ resonance in the $\eta p$ invariant mass spectra in Fig.
\ref{fig:bigger_window}. As we have already argued in Sec. 
\ref{sec:extension_higher},
the separation of the two peaks might happen at higher values of the 
incident photon
energy but should be qualitatively visible in experiment.

\vspace*{0.3cm}

At this point we would like to make some comments concerning the accuracy of our results. If one looks at the results
obtained for the $\gamma p\to \eta p$ cross section in Fig. \ref{fig:csonemeson} we can see that except in the low energy regime close to threshold, we have large discrepancies between the three options and also with experiment. The agreement can be considered just qualitatively. It is clear that some background and the contribution of the $N^*(1650)S_{11}$ might be missing and that in any case the theoretical uncertainties from different acceptable options are as big as 20-25 \% at some energies. We should not expect better agreement with experiment in the $\gamma p\to\pi^0\eta p$ reaction which requires the $\gamma p\to\eta p$ amplitude in some terms (see Fig. \ref{fig:krorumepopio}). However, as we have discussed, these terms give a small contribution to the total amplitude, since the largest contribution comes from the tree level diagram of Fig. \ref{fig:background} and its unitarization in Fig. \ref{fig:sigstardelstar}. Thus, at the end, the uncertainties in the result for the $\eta p$ invariant mass distribution, as seen in Fig. \ref{fig:pheno_pot}, are smaller than those of Fig. \ref{fig:csonemeson}. Furthermore, when one integrates over the $\eta p$ invariant mass distribution, the uncertainties in the calculation in the total cross section are relatively small, although we would not claim a precision of better than 20 \% considering all the different sources that enter the calculation. Given the complexity of the model, such an uncertainty is not easy to decrease at the present time, but it is more than acceptable for this first model of the reaction.

\subsection{The reaction $\gamma p\to \pi^0 K^0\Sigma^+$}
\begin{figure}
\includegraphics[width=6cm]{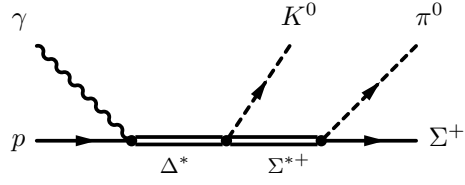}
\caption{Tree level contribution for the $\pi^0 K^0\Sigma^+$ final state.}
\label{fig:tree_level_ks}
\end{figure}
\begin{figure}
\includegraphics[width=15cm]{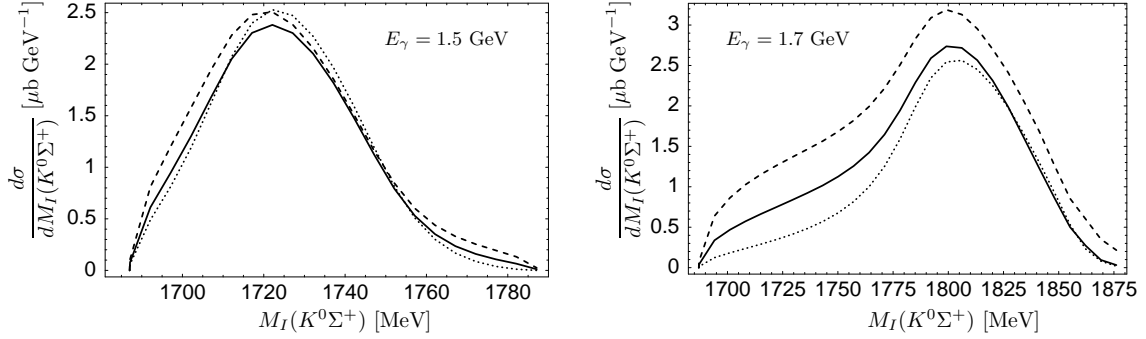}
\caption{Invariant mass spectrum for the reaction $\gamma p\to \pi^0 
K^0 \Sigma^+$ for
two photon lab energies $E_\gamma$. Solid lines: full model for the 
$N^*(1535)$. Dashed
lines: reduced model for the $N^*(1535)$. Dotted lines: Only tree 
level process from Fig.
\ref{fig:tree_level_ks}.}
\label{fig:dsdmkzerosigplus}
\end{figure}

The $\gamma p\to \pi^0 K^0 \Sigma^+$ reaction is calculated in a 
similar way as in the
last sections for the $\pi^0\eta p$ final state as the coupled 
channel formalism for the $N^*(1535)$
contains the $K^0\Sigma^+$ final state in a natural way. There is, 
however, a different tree
level diagram as displayed in Fig. \ref{fig:tree_level_ks} with the amplitude
\be
  T_{\gamma p\to\pi^0 K^0\Sigma^+}^{\rm 
BG}&=&\frac{2}{5}\frac{D+F}{2f_\pi}\;g_K
\;G_{\Sigma^*}(z'')\; G_{\Delta^*}(\sqrt{s})\vec{S}\cdot {\bf
p}_\pi\left[-ig_1'\frac{\vec{S}^\dagger\cdot {\bf 
k}}{2M}(\vec{\sigma}\times {\bf
k})\cdot\vec{\epsilon}-\vec{S}^\dagger\cdot\vec{\epsilon}\left(g_1'(k^0+\frac{{\bf
k}^2}{2M})+g_2'\sqrt{s}\;k^0\right)\right]\non
\label{tree_level_kzero_sigplus}
\ee where
\be z''=\left(s+m_{K^0}^2-2\sqrt{s}\;p_{K^0}^0\right)^{1/2}
\ee in analogy to Eq. (\ref{z3}), when expressing the amplitude in terms of the
$K^0\Sigma^+$ invariant mass. The $\Sigma^*(1385)$ propagator 
$G_{\Sigma^*}$ has been given its width as
explained below Eq. (\ref{new_sigstars}).  Note that
$T^{BG}_{\gamma p \to \pi^0K^0\Sigma^+} = t^{(6)}_{K^0
\Sigma^{*+}\Sigma^+}G_{\Sigma^*}(z'')$ with $t^{(6)}_{K^0
\Sigma^{*+}\Sigma^+}$ from Eq. (\ref{delstarksig2}) in analogy to Eq. 
(\ref{deltastarbg}).
For the contributions with rescattering, we simply
have to choose the
$(i,6)$ channel instead of the $(i,3)$ channel in $T^{(ij)}$ from Eq. 
(\ref{bse}), in the
ordering of the channels from Eq. (\ref{channels}). This means the replacement
$T^{(j3)}\to T^{(j6)}$ in Eq. (\ref{t13_contact}) and accordingly for 
the rest of the
contributions. The invariant mass distribution is obtained from a 
similar formula as Eq.
(\ref{phase_space}) with $M_i=M_{\Sigma^+}$ and is plotted in Fig.
\ref{fig:dsdmkzerosigplus}.

The tree level contribution from Fig. \ref{fig:tree_level_ks}, dotted line,
dominates the spectrum. The reaction is situated at much higher 
energies than the $\gamma p\to
\pi^0\eta p$ process and the dynamically generated $N^*(1535)$ is 
off-shell for the
entire invariant mass range. Thus, the rescattering part appears as a 
uniform background.
The tree level process shows a pronounced maximum which moves with 
the incident photon
energy. This situation is analogous to the moving peak of the 
$\Delta(1232)$ in Fig.
\ref{fig:bigger_window} and reflects the $\Sigma^*(1385)$ which is 
on-shell around the peak
position. The full and reduced model for the final state interaction (see Sec.
\ref{sec_pinetan}) differ considerably in Fig. 
\ref{fig:dsdmkzerosigplus} at $E_\gamma=1.7$ GeV (solid versus
dashed line). This is due to the fact that the model for the 
$N^*(1535)$ becomes
uncertain at these high energies as it differs from the $\pi N\to\pi 
N$ partial wave
analysis from Ref. \cite{Arndt:2003if} above $\sqrt{s}\sim 1600$ MeV. 
At lower energies, the differences are smaller.
In any case, in the energy range studied here, the dominant term is provided by the 
tree level diagram of Fig. \ref{fig:tree_level_ks}.

\begin{figure}
\includegraphics[width=8cm]{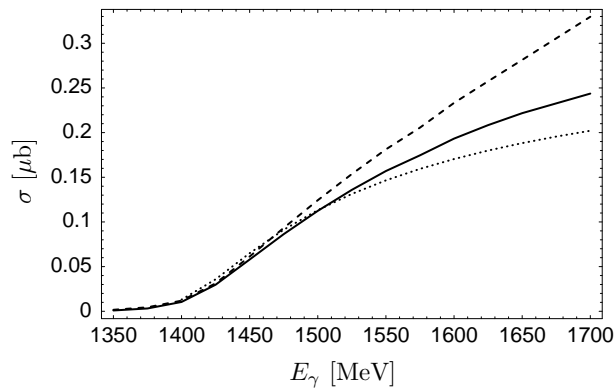}
\caption{Integrated cross section for the $\gamma p\to \pi^0 
K^0\Sigma^+$  reaction.
Solid line: Full model for the $N^*(1535)$. Dashed line: Reduced model for the
$N^*(1535)$. Dotted line: Contribution from the tree level diagram in Fig.
\ref{fig:tree_level_ks}.}
\label{fig:ksigmacross}
\end{figure}

The integrated cross section is displayed in Fig.
\ref{fig:ksigmacross}.
Comparing with Fig. \ref{fig:cross_section} it becomes 
obvious that the
production of $\pi^0 K^0 \Sigma^+$ is highly suppressed.
This is a combined effect of the $\Delta^*(1700)$ and the $N^*(1535)$
being off shell which we quantify below. Both reactions are compared at 
an energy of 150 MeV above their respective thresholds, where both cross sections
have become significantly different from zero. 
For the resulting photon energies of $E_\gamma=1202$ MeV  and 1603 MeV for the
$\pi^0\eta p$ and $\pi^0 K^0\Sigma^+$ final states, respectively, 
we obtain $\sigma_{\pi^0 \eta p}/\sigma_{\pi^0K^0\Sigma^+}=10.9$. First, this ratio is an effect
of the positive interference between the dynamically generated $N^*(1535)$ with the large
contribution of the tree level term from Fig. \ref{fig:background}.
Calculating the cross section by using this latter term only, $\sigma_{\pi^0\eta p}(1202\;{\rm MeV})$ decreases by a factor 2.0.
Second, and more important, the $\Delta^*(1700)$ propagator from Eq. (\ref{deltastarprop}) is off shell for the higher photon energy,
$|G_{\Delta^*}(1202\; {\rm MeV})|^2/|G_{\Delta^*}(1603\; {\rm MeV})|^2=5.4$. Multiplying these two factors, one obtains 10.8 which 
clarifies the origin of the factor 10.9 quoted above.

Turning the argument around, if the experiment sees a factor 10 suppression of the $\pi^0K^0\Sigma^+$ final state, compared to $\pi^0\eta p$,
this can be easily explained by the dominant role of the $\Delta^*(1700)$ found in the present study.

Other resonances beyond the $\Delta^*(1700)$ can contribute at these 
high energies, and
their omission produces uncertainties in the calculated cross 
section.  However, assuming
that we have included the relevant mechanisms in the present model, 
the suppression of
the $\pi^0 K^0\Sigma^+$ versus the $\pi^0\eta p$ final state is such a strong effect that it should be visible
in experiment.

\section{Conclusions}
In this paper we have studied the reactions $\gamma p \to \pi^0
\eta p$ and
$\gamma p \to \pi^0 K^0 \Sigma^+$ making use of a chiral unitary framework 
which considers the
interaction of mesons and baryons in coupled channels and 
dynamically generates the
$N^*(1535)$. This resonance appears  from the $s$-wave rescattering 
of $\eta N$ and
coupled channels. We have used general chiral Lagrangians  for the 
photoproduction
mechanisms and have shown that even if at tree level  the amplitudes 
for these reactions are
zero, the unitarization in coupled channels  renders the cross 
sections finite by coupling
the photon to intermediate charged meson  channels that lead to the 
$\eta p$ and
$K^0 \Sigma^+$ in the final state through multiple scattering of the 
coupled channels.

The theoretical framework has been complemented by other
ingredients, considering explicit excitation of resonances, 
whose couplings to photons are taken from experiment.

The
interaction of the meson octet with the baryon decuplet leads  to a 
set of dynamically
generated resonances, one of which has been identified with the 
$\Delta^*(1700)$. The
decay of this resonance into $\eta \Delta$ and $K\Sigma^*$, followed by the
unitarization, or in other words,  the $\Delta^*(1700)\to \pi^0 
N^*(1535)$ decay,
provides in fact the dominant contribution to the $N^*(1535)$ peak in 
the invariant mass
spectrum. A similar term provides also a tree level process which 
leads, together with
the $N^*(1535)$,  to a characteristic double hump structure in the 
$\eta p$ and $\pi^0p$
invariant mass at higher photon energies.

A virtue of this approach, concerning the $\eta p$ spectrum around
the $N^*(1535)$, and a test of the nature of this resonance 
as a dynamically
generated object, is that one can make predictions about cross 
sections for the production
of the resonance without introducing the resonance explicitly into the 
formalism; only
its components in the $(0^-, 1/2^+)$ and $(0^-, 3/2^+)$ meson-baryon 
base are what matters, together with the coupling of
the photons to these components and their interaction in a coupled 
channel formalism.
The reactions studied here also probe
decay channels of the $\Delta^*(1700)\to 
\eta\Delta(1232), \Delta^*(1700)\to K\Sigma^*(1385)$
or transitions like $\eta p\to N^*(1535)\to\eta p$ which are 
predicted by the model and not measured yet.

We have made predictions for the cross sections and for invariant mass 
distributions in the
case of the $\gamma p \to \pi^0 \eta p$ reaction. For the second 
reaction under study, $\gamma p \to
\pi^0 K^0 \Sigma^+$, we could see that in the regions
not too far from threshold of the $\gamma p \to
\pi^0 K^0
\Sigma^+$ reaction, the cross section for the latter one was much 
smaller than for the
first reaction.

The measurement of both cross sections is being performed at the 
ELSA/Bonn Laboratory
and hence the predictions are both interesting and opportune and can 
help us gain a
better insight in the nature of some resonances, particularly the
$N^*(1535)$ and the $\Delta^*(1700)$ in the present case.

\section*{Acknowledgments} We would like to acknowledge useful 
discussions with V.
Metag and M. Nanova.  This work is partly supported by DGICYT contract number
BFM2003-00856, and the E.U. EURIDICE network contract no. 
HPRN-CT-2002-00311. This
research is  part of the EU Integrated Infrastructure Initiative 
Hadron Physics Project
under  contract number RII3-CT-2004-506078.

\end{document}